\documentclass[reprint,aps,pre,floatfix,superscriptaddress,showpacs,amsfonts,
amssymb,amsmath]{revtex4-1}
\usepackage{bm}
\usepackage{epsfig,graphicx}
\usepackage{color}
\usepackage{hyperref}

\newcommand{\eq}[1]{Eq.~(\ref{#1})}
\newcommand{\eqs}[2]{Eqs.~(\ref{#1}--\ref{#2})} 
\newcommand{\Eq}[1]{Eq.~(\ref{#1})}
\newcommand{\Eqs}[2]{Eqs.~(\ref{#1}--\ref{#2})} 
\newcommand{\fig}[1]{Fig.~\ref{#1}}
\newcommand{\Fig}[1]{Fig.~\ref{#1}}

\newcommand{\secref}[1]{section~\ref{#1}}
\newcommand{\secsref}[2]{sections~\ref{#1} and~\ref{#2}}
\newcommand{\etal}{\textit{et al.}}

\newcommand{\vu}{\bm{u}}
\newcommand{\vB}{\bm{B}}

\newcommand{\delperp}{\nabla_\perp}

\newcommand{\lt}{\left}
\newcommand{\rt}{\right}

\newcommand{\bea}{\begin{eqnarray}}
\newcommand{\eea}{\end{eqnarray}}
\newcommand\be{\begin{equation}}
\newcommand\ee{\end{equation}}

\newcommand{\gmax}{\gamma_\text{max}}
\newcommand{\kmax}{\kappa_\text{max}}

\newcommand{\Lsheet}{L_\text{CS}}
\newcommand{\deltacs}{\delta_\text{CS}}

\newcommand{\DD}{\Delta'}

\newcommand{\aspratsq}{\epsilon^2}
\newcommand{\uinf}{u_\infty}
\newcommand{\normy}{\bar y}
\newcommand{\normyzero}{\bar y_0}
\newcommand{\normyzerocrit}{\bar y_{0,\rm crit}}
\newcommand{\finf}{f_\infty}
\newcommand{\fsqinf}{f^2_\infty}
\def\lambhat3/2{\hat\lambda^{3/2}}
\newcommand{\deltain}{\delta_{\rm inner}}
\newcommand{\Scrit}{S_{\rm crit}}
\renewcommand{\(}{\left(}
\renewcommand{\)}{\right)}
\newcommand{\ymyo}{\(\normy-\normyzero\)}
\renewcommand{\[}{\left[}
\renewcommand{\]}{\right]}
\newcommand{\dd}{\partial}

\begin{document}

\title{Plasmoid and 
Kelvin-Helmholtz instabilities in Sweet-Parker current sheets}
\author{N.\ F.\ Loureiro}
\affiliation{Associa\c{c}\~ao EURATOM/IST, Instituto de Plasmas e Fus\~ao 
Nuclear -- Laborat\'orio Associado, Instituto Superior T\'ecnico, Universidade 
T\'ecnica de Lisboa, 1049-001 Lisboa, 
Portugal}
\author{A.\ A.\ Schekochihin}
\affiliation{Rudolf Peierls Centre for Theoretical Physics, University of 
Oxford, Oxford OX1 3NP, UK}
\author{D.\ A.\ Uzdensky}
\affiliation{Center for Integrated Plasma Studies, Physics Department, 
University of Colorado, Boulder CO 80309, USA} 
\date{\today}

\begin{abstract}
A 2D linear theory of the instability of Sweet-Parker (SP) 
current sheets is developed in the framework of Reduced MHD. 
A local analysis is performed taking into account the dependence 
of a generic equilibrium profile on the outflow coordinate.
The plasmoid instability [Loureiro \etal, Phys. Plasmas {\bf 14},
 100703 (2007)] is recovered, i.e., 
current sheets are unstable to 
the formation of a large-wave-number chain of plasmoids
 ($k_{\rm max}\Lsheet \sim S^{3/8}$, where $k_{\rm max}$ is the 
wave-number of fastest growing mode, 
$S=\Lsheet V_A/\eta$ is the Lundquist number, 
$\Lsheet$ is the length of the sheet, $V_A$ is the Alfv\'en speed and $\eta$ 
is the plasma resistivity), which grows super-Alfv\'enically fast
 ($\gmax\tau_A\sim S^{1/4}$, where $\gmax$ is the maximum growth rate, and $\tau_A=\Lsheet/V_A$).
For typical background profiles, the growth rate and the wave-number 
are found to {\it increase} in the outflow direction.
This is due to the presence of another mode, the Kelvin-Helmholtz (KH) instability, which is 
triggered at the periphery of the layer, 
where the outflow velocity  exceeds the Alfv\'en speed associated with the upstream 
magnetic field.
The KH instability grows even faster than 
the plasmoid instability, $\gmax \tau_A \sim  k_{\rm max} \Lsheet\sim S^{1/2}$.
The effect of viscosity ($\nu$) on the plasmoid instability is also addressed. 
In the limit of large magnetic  Prandtl numbers, 
$Pm=\nu/\eta$, it is found that 
$\gmax\sim S^{1/4}Pm^{-5/8}$ and $k_{\rm max} \Lsheet\sim S^{3/8}Pm^{-3/16}$, leading to the 
prediction that 
the critical Lundquist number for plasmoid instability 
in the $Pm\gg1$ regime 
is $\Scrit\sim 10^4Pm^{1/2}$. These results are verified via direct numerical simulation of the 
linearized equations, using a new, analytical 2D SP equilibrium solution.
\end{abstract}


\maketitle
\section{Introduction}
\label{intro}
Magnetic reconnection~\cite{biskamp_magnetic_2000,priest_magnetic_2000,
zweibel_magnetic_2009, yamada_magnetic_2010} is a ubiquitous plasma physics phenomenon, 
characterized by the rapid reconfiguration of the 
magnetic-field topology. 
Solar flares~\cite{shibata_solar_2011} and magnetospheric substorms~\cite{oieroset_situ_2001} 
are two prominent examples 
of events where reconnection plays a key role.
Plasma dynamics in many laboratory experiments
is also critically determined by 
magnetic reconnection; examples are the 
sawtooth~\cite{hastie_sawtooth_1997} and the
tearing instabilities~\cite{furth_finite-resistivity_1963,rutherford_nonlinear_1973} in 
magnetic-confinement-fusion devices,
or the reconnection of high-energy-density, 
laser-produced plasma bubbles~\cite{nilson_magnetic_2006,willingale_proton_2010, 
dong_plasmoid_2012}.

Along with fast reconnection rates, many observations~\cite{lin_plasmoids_2008} of
magnetic reconnection phenomena display 
one intriguing feature: 
the formation, and subsequent ejection from the current sheet,
of coherent secondary 
structures, often referred to as plasmoids (also known as blobs, 
flux ropes or secondary magnetic islands).
There is abundant direct evidence 
for the presence of these structures in the Earth's 
magnetotail~\cite{zong_cluster_2004, eastwood_observations_2005,chen_observation_2008}
and in solar flares~\cite{lin_direct_2005,ciaravella_current_2008,
bemporad_spectroscopic_2008,lin_investigation_2009, nishizuka_multiple_2010, 
takasao_simultaneous_2012}. 
In magnetic confinement fusion devices, plasmoid generation seems 
to be less certain, 
though there are reports of the observation of secondary magnetic structures correlated to 
 $m/n=1/1$ and $m/n=2/1$ magnetic islands on the 
TEXTOR~\cite{donne_evidence_2005,liang_observations_2007} and JET~\cite{salzedas_secondary_2011} 
tokamaks.
On TEXTOR, high-resolution measurements of electron temperature 
fluctuations show structures which hint at
plasmoid formation during sawtooth crashes~\cite{park_observation_2006,park_comparison_2006,
munsat_letter:_2007}. 
Finally, recent laser-plasma experiments where reconnection is conjectured to occur 
also show evidence for plasmoid formation~\cite{willingale_proton_2010,dong_plasmoid_2012}. 

Direct numerical simulations of reconnection processes
concur with observations in displaying ubiquitous evidence
for plasmoid formation.
Plasmoids have been reported in numerical simulations using various physical 
models, ranging from 
kinetic~\cite{daughton_fully_2006,drake_formation_2006,
karimabadi_multi-scale_2007,daughton_transition_2009, daughton_influence_2009} to 
Hall-MHD~\cite{shepherd_comparison_2010,huang_onset_2011} and to 
single fluid MHD~\cite{park_reconnection_1984,steinolfson_nonlinear_1984,lee_multiple_1986,
biskamp_magnetic_1986,jin_twodimensional_1991,loureiro_x-point_2005-1,
lapenta_self-feeding_2008,samtaney_formation_2009,loureiro_turbulent_2009,
bhattacharjee_fast_2009, cassak_scaling_2009-1, huang_scaling_2010, skender_instability_2010, 
loureiro_magnetic_2012}.
Plasmoid formation has also been reported in numerical simulations of reconnection 
in relativistic plasmas, both 
resistive~\cite{komissarov_tearing_2007} and 
kinetic~\cite{zenitani_role_2008,liu_particle_2011,sironi_acceleration_2011,cerutti_beaming_2012}.
Numerical studies tailored to address specific reconnection
contexts such as the solar corona~\cite{riley_``bursty_2007,barta_dynamics_2008,
bettarini_spontaneous_2010,barta_spontaneous_2010}, 
the Earth's magnetotail~\cite{karimabadi_magnetic_1999, jin_2.5_2001}, magnetic young stellar 
objects~\cite{goodson_jets_1999,goodson_jets_1999-1,uzdensky_magnetic_2004},
fusion experiments~\cite{park_reconnection_1984,biskamp_magnetic_1986} and 
laser-plasma interactions~\cite{fox_fast_2011},
though different from each other in a number of details, 
again all appear to agree on the 
basic fact that plasmoids are generated in reconnecting current sheets.

The plasmoid dynamics inferred from 
observations and seen in numerical simulations strongly suggest the 
very \textit{opposite} 
of the laminar, steady-state reconnection scenarios that have dominated the 
field for much of its history [the  Sweet-Parker (SP)~\cite{parker_sweets_1957,sweet_neutral_1958} 
and the 
Petschek~\cite{petschek_magnetic_1964} models,
and, more recently, the Hall reconnection
paradigm~\cite{birn_geospace_2001}]. 
Magnetic reconnection in the presence of plasmoids appears to 
be a highly time-dependent, bursty process, which can only be described in 
a statistical manner~\cite{shibata_plasmoid-induced-reconnection_2001,barta_spontaneous_2010, 
fermo_statistical_2010,uzdensky_fast_2010, fermo_comparison_2011, loureiro_magnetic_2012}.
Furthermore, in addition to their key role in setting the reconnection 
rate in both laminar~\cite{shibata_plasmoid-induced-reconnection_2001, lapenta_self-feeding_2008, 
daughton_transition_2009, 
daughton_influence_2009,bhattacharjee_fast_2009,cassak_scaling_2009-1,huang_scaling_2010,
uzdensky_fast_2010,huang_onset_2011,loureiro_magnetic_2012} and 
turbulent~\cite{loureiro_turbulent_2009,skender_instability_2010} plasmas, 
there is numerical and observational evidence that 
plasmoids may be critical in explaining electron acceleration in reconnection 
sites~\cite{drake_electron_2006, chen_observation_2008, oka_island_2010, oka_electron_2010}.

In a previous paper~\cite{loureiro_instability_2007} (henceforth referred to as Paper I)
we attempted to understand the origin of plasmoid formation in reconnection sites by analysing 
the linear stability of 
large-aspect-ratio, SP current sheets. 
These were found to be violently unstable to the formation of
plasmoid chains, the  fastest growing wave number scaling as $k_{\rm max} \Lsheet \sim S^{3/8}$, 
with corresponding growth rate $\gmax \tau_A\sim S^{1/4}$, where $\Lsheet$ is the length of the 
current layer, $\tau_A=\Lsheet/V_A$ is the Alfv\'en time ($V_A$ is the Alfv\'en speed) and 
$S$ is the Lundquist number, $S=\Lsheet V_A/\eta$, where $\eta$ is the magnetic diffusivity.
Since $S\gg 1$ in most applications of interest, this theory predicts the formation of 
multidinous plasmoids growing super-Alfv\'enically;
the immediate implication is that stable reconnecting current sheets at large 
values of the Lundquist number cannot exist. 
These results have since been confirmed in direct 
numerical simulations~\cite{samtaney_formation_2009, ni_linear_2010}, and extended to 
account for the effect of 
a finite component of the magnetic field perpendicular to the reconnection 
plane~\cite{baalrud_reduced_2012} 
and into the two-fluid regime~\cite{baalrud_hall_2011}.

The analysis of Paper I considered a very simplified model background equilibrium, intended 
to retain only what we viewed as the most important features of a SP
current sheet: a reconnecting magnetic field, $\bm B_{eq}=(0,B_y(x))$ ($x$ is the inflow direction, 
$y$ the outflow direction) and an incompressible flow 
defined by the stream function $\phi_{eq}=\Gamma_0xy$, where 
$\Gamma_0=V_A/\Lsheet$
is the flow shearing rate. 
The analytical derivation in Paper I did not, therefore, take into account
potentially important effects, such as  
the variation of the reconnecting magnetic field and of the outflow speed along the layer 
(i.e., along the $y$-direction in 
our chosen geometry), or the reconnected magnetic field.

In this work, we generalize the results of Paper I
to a more 
realistic, two-dimensional model of the current sheet.
Using an approach in the spirit of WKB theory (justified by the expectation that 
the most unstable wave-number will be very large, $k_{\rm max}\Lsheet\gg1$), we derive the 
dispersion relation for the plasmoid instability as
a slow function of the position along the sheet, $y_0$.
We find that the scalings of the maximum growth rate ($\gmax$) 
and wave-number ($k_{\rm max}$) derived in Paper I 
hold true in a central, finite-sized patch of  the current sheet; however, the growth rate 
and wave number are now parametrized nontrivially by $y_0$.
Surprisingly, we also discover that for a generic background equilibrium configuration, 
the maximum growth rate and wave-number of the 
instability \textit{increase} with $y_0$ (i.e., outwards).
As we show in this paper, a special point exists, $y_{0,\rm crit}$, beyond which the 
assumptions invoked in our calculation break down.
This is the Alfv\'en Mach point of the system, 
where the magnitude of the outflow velocity (an increasing function of $y_0$) 
becomes equal to the value of the 
Alfv\'en speed based on the upstream magnetic field (a decreasing function of $y_0$).
Beyond that point, the current sheet becomes unstable to 
a different mode: the Kelvin-Helmholtz (KH) instability, whose growth rate and 
wave-number dependence we also derive analytically.

The other main result of this paper is the study of the effect of a large viscosity $\nu$ 
(parametrized by the magnetic Prandtl number $Pm = \nu/\eta\gg 1$) 
on the plasmoid instability. The large-Prandtl-number regime is pertinent to various astrophysical 
applications, e.g., the interstellar medium~\cite{kulsrud_spectrum_1992}, and to fusion 
plasmas~\cite{park_reconnection_1984}, and so it is important to understand 
how large $Pm$ affects plasmoid formation and dynamics.

Our analytical results are complemented with a direct numerical solution of the full set of 
linearized equations.

This paper is organized as follows. 
In~\secref{sec:heuristic}, we present a heuristic derivation of our main results.
A more rigorous approach to the problem begins in \secref{sec:prob_setup}, where the 
equations to be solved are laid out and the expected 
properties of the background equilibrium 
are discussed (a more quantitative discussion of the constraints that the background 
equilibrium should satisfy can be found in Appendix~\ref{equilibrium}, where an analytical 
2D SP-like current-sheet equilibrium is obtained). 
The core of the analytical calculation is presented in \secref{sec:lin_theory}.
The KH instability of the current sheet is derived in~\secref{KH}.
Results of the direct numerical solution of the linear equations
are presented in~\secref{numerics}. The effect of viscosity 
on the instability is addressed in~\secref{sec:viscosity}.
Finally, a discussion of the results and conclusions can be found in~\secref{sec:conclusions}.
\section{Heuristic Derivation}
\label{sec:heuristic}
In this section, we show how the main results of this paper can be derived 
in a simple (albeit non-rigorous) way. 
A reader uninterested in the formal mathematical 
details can skip to \secref{numerics} after this section.

\subsection{Plasmoid instability}
\label{heur_plasmoids}
The fastest growth rate of the plasmoid instability can be obtained 
from the usual tearing mode formulae as follows~\cite{bhattacharjee_fast_2009}.

In the small $\DD$ limit, where $\DD$ is the usual tearing mode instability parameter,  
the standard (FKR) tearing mode dispersion relation is~\cite{furth_finite-resistivity_1963} 
\be
\label{gamma_FKR}
\gamma\sim \tau_H^{-2/5} \tau_\eta^{-3/5} (\DD a)^{4/5},
\ee
where $a$ is the characteristic equilibrium magnetic field length scale, 
$\tau_\eta=a^2/\eta$ is the resistive diffusion time and $\tau_H = 1/kB_0$ is the 
hydrodynamic time. 
In the opposite limit of large $\DD$~\cite{coppi_1976},
\be
\label{gamma_Coppi}
\gamma\sim \tau_H^{-2/3} \tau_\eta^{-1/3}.
\ee
In both cases, the width of the resistive (or inner) boundary layer is
\be
\label{delta_FKR}
\deltain/a\sim \(\gamma \tau_H^2 \tau_\eta^{-1}\)^{1/4}.
\ee

To find the fastest growing mode, let us assume a simple Harris-sheet equilibrium, 
$B_y = B_0\tanh(x/a)$; 
then, for $ka\ll 1$, as we expect to be the case for $k=k_{\rm max}$, 
we have $\DD a \sim 1/ka$. Substituting this expression in \eq{gamma_FKR}, we 
find that it yields $\gamma\propto k^{-2/5}$, whereas from \eq{gamma_Coppi} we have
$\gamma\sim k^{2/3}$. 
Approximate expressions
for the largest growth rate and corresponding wave number can therefore be found 
by balancing Eqs.~(\ref{gamma_FKR}) and (\ref{gamma_Coppi}).
This gives
\bea
k_{\rm max} a &\sim& (a/B_0)^{1/4} \tau_\eta^{-1/4},\\
\gmax &\sim& (a/B_0)^{-1/2} \tau_{\eta}^{-1/2}.
\eea
The corresponding inner-layer width is
\be
\label{delta_FKR_max}
\deltain/a\sim (a B_0 /\eta)^{-1/4}.
\ee

In order to apply these scalings to a SP current sheet, we
rescale the equilibrium length scale $a$ to the sheet thickness:
\be
\label{eq_rescale}
a\equiv \deltacs \sim \Lsheet S^{-1/2}.
\ee
Noting that the plasma outflow speed in a SP current sheet is $V_A=B_0$,
we obtain
\bea
\label{kmax_plasmoid}
k_{\rm max} \Lsheet &\sim& S^{3/8},\\
\label{gmax_plasmoid}
\gmax \tau_A &\sim& S^{1/4},\\
\label{deltain_plasmoid}
\deltain/\deltacs&\sim& S^{-1/8},
\eea
where $\tau_A=\Lsheet/V_A$. 
These predictions are in agreement with the results of Paper I~\cite{loureiro_instability_2007}.

We can now use these results to estimate the critical value of the Lundquist number, $\Scrit$, 
below which 
we expect SP current sheets to be stable. The underlying reasoning is that for the 
plasmoid instability to be triggered we must have 
$\gmax\tau_A\gg 1,~k_{\rm max}\Lsheet \gg 1$ and $\deltain/\deltacs \ll 1$, i.e., the instability 
has to grow faster than the characteristic outflow time and it has to fit inside the current sheet, 
both along (thus the restriction on $k_{\rm max}$) and across (thus the restriction on $\deltain$).
Of all these conditions, the most stringent is that on the width of the inner layer, since it 
bears the weakest $S$ dependence. 
Therefore, if we require (non-rigorously!) that $\deltain/\deltacs \sim 1/3$ at the very most,
then \eq{deltain_plasmoid} would yield 
$\Scrit\sim 10^4$. This is consistent with numerical 
simulations~\cite{lee_multiple_1986,biskamp_magnetic_1986, loureiro_x-point_2005-1, 
samtaney_formation_2009}.

\subsection{Plasmoid instability at large Pm}
\label{heur_largePm}
One limitation of  Paper I was
that plasma viscosity was neglected. 
At low values of the magnetic Prandtl number, 
$Pm=\nu/\eta$ [relevant to the interiors of stars and planets, or liquid metal 
laboratory dynamos, for example [see~\cite{schekochihin_fluctuation_07} and references therein)], 
the presence of viscosity should not
change our results substantially.
In contrast, for $Pm\gg1$, as is often found in fusion plasmas~\cite{park_reconnection_1984}, 
warm interstellar and intracluster media~\cite{kulsrud_spectrum_1992,brandenburg_astrophysical_2005}, 
etc., 
both the SP scalings and the tearing and kink modes
scalings change~\cite{porcelli_viscous_1987}, and so, therefore, will the plasmoid instability.

Let us work out  the plasmoid scalings in the large-$Pm$ 
limit in a similar way to that just presented for the invisicid case.
Instead of the FKR~\cite{furth_finite-resistivity_1963} and Coppi {\it et al.}~\cite{coppi_1976} 
results, we now use the corresponding formulae
valid for $Pm\gg 1$, i.e., the
the so-called ``visco-tearing'' (low $\DD$) and ``visco-resistive kink'' (large $\DD$) derived 
by Porcelli~\cite{porcelli_viscous_1987}. 

At low $\DD$ (the visco-tearing mode), we have
\be
\label{Porc_VT}
\gamma\sim \tau_H^{-1/3}\tau_\eta^{-5/6}\tau_\nu^{1/6}\DD a,
\ee
where  $\tau_\nu=a^2/\nu$ is the viscous diffusion time.
At $\DD\rightarrow\infty$ (the visco-resistive kink), the growth rate is
\be
\label{Porc_VRK}
\gamma\sim \tau_H^{-2/3}\tau_\eta^{-2/3}\tau_\nu^{1/3}.
\ee
The corresponding inner-layer width is
\be
\label{deltain_visc}
\deltain/a\sim \[\tau_H^2/\(\tau_\eta \tau_\nu\)\]^{1/6}.
\ee

As before, let us assume that $\DD a \sim 1/ka$ for $k\sim k_{\rm max}$. Then
we find from~\eq{Porc_VT} that $\gamma\propto k^{-2/3}$, 
whereas~\eq{Porc_VRK} yields $\gamma\propto k^{2/3}$. Scalings
for the fastest growing mode can thus again be found by balancing Eqs. (\ref{Porc_VT}) 
and (\ref{Porc_VRK}).
The result is
\bea
k_{\rm max} a &\sim& (a/B_0)^{1/4} \tau_\eta^{-1/8} \tau_{\nu}^{-1/8},\\
\gmax &\sim& (a/B_0)^{-1/2} \tau_{\eta}^{-3/4} \tau_\nu ^{1/4},\\
\deltain/a &\sim& (a/B_0)^{1/4} \tau_\eta^{-1/8} \tau_{\nu}^{-1/8}.
\eea

We now repeat the previous procedure of rescaling the equilibrium length scale $a$, 
this time using the results obtained by Park {\it et al.}~\cite{park_reconnection_1984} for the SP model 
in the limit $Pm\gg1$:
\be
a\equiv \deltacs \sim \Lsheet S^{-1/2} Pm^{1/4}.
\ee
This gives
\bea
\label{kmax_largePm}
k_{\rm max} \Lsheet &\sim& S^{3/8} Pm^{-3/16},\\
\label{gmax_largePm}
\gmax \tau_A &\sim& S^{1/4} Pm^{-5/8},\\
\label{deltain_largePm}
\deltain/\deltacs &\sim& S^{-1/8} Pm ^{1/16}.
\eea
We shall find in~\secref{numerics} that these scalings indeed agree very well 
with the results of a  direct numerical integration of the 
linearized equations.
We thus find that the dependence of $\gmax$ and $k_{\rm max}$ on $S$ remains 
unchanged at large $Pm$.
However, viscosity damps the instability, and decreases the wave-number and the growth rate
of the fastest growing mode, while slightly thickening the inner layer.

An important question is how the critical Lundquist number for the onset of the 
current sheet instability, $\Scrit$, scales with the magnetic Prandtl number. 
Although the expressions 
above are formally only valid in the limit $S\gg \Scrit,~Pm\gg1$, we can 
use them to obtain a rough estimate of this dependence.
Since the instability requires $\deltain/\deltacs\ll1$, we may 
again demand that this be at most $1/3$ and use
\eq{deltain_largePm} to obtain
\be
\Scrit\sim 10^4 Pm^{1/2}.
\ee

To see the consistency of this result, note that the same dependence of $\Scrit$ on $Pm$
can be obtained by either looking for the minimum wavenumber
that will fit inside the current sheet, $k_{\rm max}\Lsheet\sim 1$, or by requiring that 
the growth rate is comparable to the flow shear rate, $\gmax \Lsheet/u_{\rm out}\sim1$ 
(note that for $Pm\gg 1$, 
$u_{\rm out}\sim V_A Pm^{-1/2}$~\cite{park_reconnection_1984}). 
In both cases, \eq{kmax_largePm} and 
\eq{gmax_largePm} yield $\Scrit\sim Pm^{1/2}$. 
This result is a specific prediction, which can in principle be checked via direct 
numerical simulations of the current-sheet instability in the large-Prandtl-number regime.

\subsection{Kelvin-Helmholtz instability}
\label{heur_KH}
\begin{figure}
\includegraphics[width=0.48\textwidth]{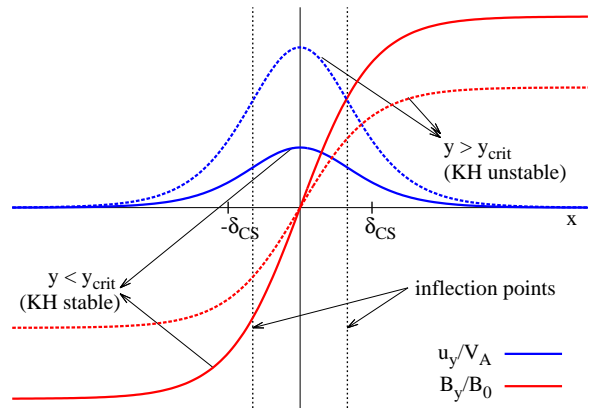}
\caption{Cartoon illustrating KH-stable and KH-unstable parts of an idealized SP current sheet.
The outflow profile $u_y$ is depicted in blue; the upstream magnetic field $B_y$ in red.
The vertical dashed lines mark the position of KH-unstable layers (i.e., the inflection points 
of the outflow profile, where $d^2u_y/dx^2=0$).
The full lines show the profiles below the Alfv\'en Mach point ($y<y_{\rm crit}$); in this 
case the magnitude of $B_y$ at $x\sim \pm \deltacs$ exceeds that of $u_y$ at $x=0$, and thus the 
magnetic field is sufficiently strong to stabilise the current sheet against the KH instability.
The opposite case of profiles above the Alfv\'en Mach point ( $y>y_{\rm crit}$) 
is shown by the dashed lines; $B_y$ at the 
inflow is now weaker than $u_y$ at the center of the sheet, which is, therefore, KH-unstable. 
}
\label{fig:KH_cartoon}
\end{figure}

The velocity outflow profile of a Sweet-Parker reconnection configuration is 
such that it is maximum at the 
midplane ($x=0$) of the current sheet, and decays to zero away from it. Thus, there are two 
parallel shear layers, with two corresponding inflection points 
of the outflow,  at $x\sim\pm\deltacs$.
The shear (in the $x$-direction) of this flow profile can be estimated as:
\be
\label{KH_flowprof}
\frac{d u_y}{dx}\sim \frac{V_A}{\deltacs}\frac{y}{\Lsheet}.
\ee
Each of these layers would be Kelvin-Helmholtz unstable were it not for the 
stabilizing effect of the upstream magnetic field, $B_y$~\cite{chandrasekhar_hydrodynamic_1961}:
as is well known, a magnetic field that is coplanar with the flow profile will stabilize the 
KH instability as long as $|B_y| > |u_y|$.
In the case of a SP current sheet, the upstream magnetic field $B_y$ is not constant along the 
sheet; in particular, its 
magnitude {\it decreases} in the $y$-direction; 
a simple model for it is~\cite{syrovatskii_formation_1971}
\be
\label{KH_syro}
B_y = B_0\sqrt{1-y^2/\Lsheet^2}.
\ee
It is thus possible that there exists a location $y_{\rm crit}/\Lsheet \sim 1$ along the sheet where 
the (decreasing) strength of $B_y$ matches the (increasing) magnitude of the ouflow, 
$u_y\sim V_A~y/\Lsheet $.
This is the Alfv\'en Mach point of the system;
for $y>y_{\rm crit}$, the magnetic field is no longer able to stabilize the KH mode.
\Fig{fig:KH_cartoon} provides a schematic illustration of both types (KH stable and unstable) 
of configuration.

A rigorous derivation of this instability is presented in~\secref{KH}.
Here we show how the basic scalings for the fastest growing mode and corresponding wave number 
can be obtained heuristically. From the standard  theory of the KH 
instability~\cite{chandrasekhar_hydrodynamic_1961}, it is known that
\be
\gamma_{\rm max}^{\rm KH} \sim \frac{d u_y}{dx}\sim\frac{V_A}{a},
\qquad k_{\rm max}^{\rm KH}a\sim 1,
\ee
where $a$ is the characteristic scale length of the sheared flow profile $u_y(x)$. 
As above, these estimates can be applied to the SP-sheet by simply setting
$a\rightarrow\deltacs\sim \Lsheet S^{-1/2}$, implying that $k_{\rm max}^{\rm KH}\sim 1/\deltacs$. 
Thus,
\be
\label{heur_KH_scalings}
\gamma_{\rm max}^{\rm KH}\tau_A\sim S^{1/2}, \qquad k_{\rm max}^{\rm KH}\Lsheet \sim S^{1/2}.
\ee
Note that this growth rate is even larger (i.e., has a steeper positive scaling with $S$) than 
that of the plasmoid instability, \eq{gmax_plasmoid}.

The KH instability scalings in the large $Pm$ limit are easily obtained in a similar way, 
using the following modifications of the SP relations derived in~\cite{park_reconnection_1984}: 
$u_{\rm out}\sim V_A Pm^{-1/2}$, $a\rightarrow\deltacs\sim \Lsheet S^{-1/2}Pm^{1/4}$. Then we have
\be
\label{heur_KH_scalings_largePm}
\gamma_{\rm max}^{\rm KH}\tau_A\sim S^{1/2}Pm^{-3/4}, 
\qquad k_{\rm max}^{\rm KH}\Lsheet \sim S^{1/2}Pm^{-1/4}.
\ee

\section{Problem Setup}
\label{sec:prob_setup}
In this section we proceed to make some of the above discussion more rigorous 
and quantitative.
 
We solve the 2D reduced-MHD equations~\cite{strauss_nonlinear_1976},
\bea
\label{RMHD_vort}
&&\dd_t\delperp^2\phi + \lt\{\phi,\delperp^2\phi\rt\} =
\lt\{\psi,\delperp^2\psi\rt\}+\nu\delperp^4\phi,\\
\label{RMHD_psi}
&&\dd_t\psi + \lt\{\phi,\psi\rt\} = 
\eta\delperp^2\psi - E_0.
\eea
Here $\phi$ and $\psi$ are the stream and flux functions 
of the in-plane velocity and magnetic field, respectively, so 
$\vu=(-\dd_y\phi,\dd_x\phi)$, $\vB=(-\dd_y\psi,\dd_x\psi)$;
the magnetic field is measured in velocity units;
Poisson brackets are denoted by
$\lt\{\phi,\psi\rt\}=\dd_x\phi\dd_y\psi-\dd_y\phi\dd_x\psi$; 
$\eta$ and $\nu$ denote the plasma resistivity and viscosity, respectively;
$E_0$ represents an externally applied electric field, required 
to sustain an equilibrium in the presence of finite resistivity.
In the analytical calculation that follows, we will assume that the magnetic 
Prandtl number is small, 
$Pm =\nu/\eta\ll1$ and therefore neglect the effect of 
viscosity on the linear instability. The case of large $Pm$ will be studied numerically 
in~\secref{sec:viscosity}.

We are interested in analysing the linear stability of an SP-like current sheet, 
whose inverse aspect ratio, $\epsilon$, is predicted by the SP model to scale as
\be
\label{epsilon_def}
\epsilon=\deltacs/\Lsheet\sim S^{-1/2}\ll 1.
\ee
Therefore, in the vicinity of a general point along the current sheet, 
$y=y_0$, and provided that $y_0$ is not too close to either of the ends
of the sheet, it is reasonable to expand the equilibrium 
magnetic flux and stream functions in a power 
series~\cite{biskamp_nonlinear_1993}:
\bea
\label{psi_eq}
\psi_{eq}(x,y)|_{y=y_0}&=&\sum_{n=0}^{\infty}
\frac{1}{n!}
\(\frac{y-y_0}{\Lsheet}\)^n\psi_n(x,y_0),\\
\label{phi_eq}
\phi_{eq}(x,y)|_{y=y_0}&=&\sum_{n=0}^{\infty}
\frac{1}{n!}
\(\frac{y-y_0}{\Lsheet}\)^n\phi_n(x,y_0).
\eea
The functions $\psi_n(x,y_0),~\phi_n(x,y_0)$ can in 
principle be found either by substituting these expansions back 
into \eqs{RMHD_vort}{RMHD_psi} and solving the equilibrium problem order by 
order in $(y-y_0)/\Lsheet$, or by Taylor expanding a known 
equilibrium solution, $\psi_{eq},~\phi_{eq}$ around the point 
$y=y_0$.
Neither of these procedures is straighforward: the first option implies either 
truncating the expansion at some arbitrary order or guessing one of the 
equilibrium functions~\cite{biskamp_nonlinear_1993} to solve the closure problem; 
the second requires an exact analytical solution.
To the best of our knowledge, a 2D analytical solution to this problem
that would  capture all the essential features of a resistive SP-like 
current sheet equilibrium  has never been derived~\footnote{
One should note that, starting with the 
Syrovatskii solution~\cite{syrovatskii_formation_1971}, the literature 
is rich in \textit{ideal} current sheet equilibrium solutions but, by 
definition, those 
cannot describe reconnection or the internal structure of the current sheet 
(see Ref.~[\onlinecite{priest_magnetic_2000}] for 
an extensive literature review on this subject). 
Families of solutions of the resistive equilibrium problem were derived 
by Craig and Henton~\cite{craig_exact_1995}, but none of these models 
the conventional 
SP-like configuration that we wish to treat here. 
Biskamp~\cite{biskamp_nonlinear_1993}, employing the same 
power-series expansion as used here, avoids the closure problem by
introducing an \textit{ad hoc} guess for the functional form of
the unreconnected magnetic field profile. 
In Appendix~\ref{equilibrium}, we obtain an analytic equilibrium solution 
that is a 2D generalization of Biskamp's.
}.
Fortunately, however, we shall find 
in the following sections that, actually deriving the linear instability
requires very little information about the equilibrium profiles, 
and the problem can be solved for general functions $\psi_n, ~\phi_n$ provided 
that the following key assumptions hold:
\begin{itemize}
\item the background equilibrium observes the expected symmetries, i.e.,
at the center of the sheet
\bea
\begin{array}{cc}
\psi_{2n+1}(y_0=0)=0,&
\quad\phi_{2n}(y_0=0)=0;
\end{array}
\eea

\item for $|x|\gg1$, the incoming flow $u_x$ and the reconnecting magnetic 
field $B_y$ approach constant (in $x$) values 
(which can, however, be functions of $y_0$).
\end{itemize}
Further general properties of the equilibrium that will be needed in 
our calculation are derived in Appendix~\ref{equilibrium}.

Note that, in Paper I, we considered only the case $y_0=0$. Furthermore,
we adopted a very simplified description of an 
equilibrium current sheet in which only $\phi_1$ and $\psi_0$ 
were non-zero, with 
$\phi_1=\Gamma_0xy,~\psi_0=\psi_0(x)$,
i.e., the flow had no vorticity $\omega_z=\delperp^2\phi_1=0$,
and the reconnected field was ignored, $B_x=-\dd\psi/\dd y=0$.
In this paper, we drop those model assumptions and consider an arbitrary, 
two-dimensional current-sheet equilibrium. \\

\subsection{Normalizations}
\label{sec:norms}
We introduce the following normalizations, motivated by the 
SP scalings:
\bea
\label{normalizations}
\begin{array}{ll}
\phi'_0=\Gamma_0 y_0 v(x); &
\quad \psi'_0=B_0 f(x);\\
\phi_1=-\Gamma_0 \Lsheet \deltacs u(x); &
\quad \psi_1=-\Gamma_0 y_0 \deltacs g(x); \\
\phi_2=-\Gamma_0 y_0 \deltacs w(x); &
\quad \psi_2=-\Gamma_0 \Lsheet \deltacs h(x); 
\end{array}
\eea
\\
where 
\be
\label{deltacs_def}
\deltacs=(\eta/\Gamma_0)^{1/2}
\ee
and $\Gamma_0=B_0/\Lsheet$ (note that in our units $B_0=V_A$).
We also normalise time and lengths as follows:
\be
\label{time_lengths_norm}
\begin{array}{lcr}
t\Gamma_0=\tau; &\quad x/\deltacs=\xi; &\quad y/\Lsheet=\normy.
\end{array}
\ee
Under these normalizations, the magnetic and velocity fields 
obtained from the power-series expansions~(\ref{psi_eq}--\ref{phi_eq})
keeping only up to first-order corrections in $\normy-\normyzero$ are:
\bea
\label{By_norm}
B_y/B_0 &=&f(\xi)-(\normy-\normyzero)\normyzero g'(\xi) \mbox{ (reconnecting)},\\
\label{Bx_norm}
B_x/B_0 &=&\epsilon\[\normyzero g(\xi)+\(\normy-\normyzero\) h(\xi)\] \mbox{ (reconnected)},
\eea
and
\bea
\label{u_field_norms}
u_y/V_A&=&\normyzero v(\xi)-\(\normy-\normyzero\)u'(\xi) \mbox{ (outflow)},\\
u_x/V_A&=&\epsilon\[u(\xi)-\(\normy-\normyzero\)\normyzero w(\xi)\] \mbox{ (inflow)}.
\eea
It is clear from these expressions what the physical significance 
of the functions $f,~g,~h,~v,~u,~w$ is.
The physical units and presumed magnitudes of these fields have been 
absorbed into the normalizations, so these functions are 
all order-unity dimensionless quantities.
They can have parametric dependence on $\normyzero$, but note that the 
presumed 
lowest-order linear dependence of the reconnected magnetic field and of 
the outflow on $\normyzero$ are explicitly included in the normalizations~\eq{normalizations}.
\subsection{Linearized Equations}
Let us consider small perturbations to a generic equilibrium, 
$\psi=\psi_{eq}+\delta\psi(x,y,t), ~\phi=\phi_{eq}+\delta\phi(x,y,t)$, and 
linearize the RMHD \eqs{RMHD_vort}{RMHD_psi} 
using the expansions~(\ref{psi_eq}--\ref{phi_eq}) for the equilibrium profiles,
keeping terms up to first order in $\ymyo$. We obtain
\begin{widetext}
\begin{align}
\label{lin_ohm}
\frac{\dd \delta\psi}{\dd \tau}+
\[\normyzero v(\xi)-\ymyo u'(\xi)\]
\frac{\dd \delta \psi}{\dd \normy}
+\[u(\xi)+\ymyo \normyzero w(\xi)\]\frac{\dd \delta \psi}{\dd \xi}
\nonumber\\
-\[\normyzero g(\xi)+\ymyo h(\xi)\]
\frac{\dd \delta\phi}{\dd \xi}
-\[f(\xi)-\ymyo \normyzero g'(\xi)\]\frac{\dd \delta \phi}{\dd \normy}=
{}&\(\frac{\dd^2}{\dd \xi^2}+\aspratsq\frac{\dd^2}{\dd\normy^2}\)\delta\psi
+\mathcal O\(\ymyo^2\),
\end{align}
\begin{align}
\label{lin_vort}
\left\{\frac{\dd}{\dd \tau}+
\[\normyzero v(\xi)-\ymyo u'(\xi)\]\frac{\dd}{\dd\normy}+
\[u(\xi)+\ymyo \normyzero w(\xi)\]\frac{\dd}{\dd\xi}\right\}
\(\frac{\dd^2}{\dd \xi^2}+
\aspratsq \frac{\dd^2}{\dd\normy^2}\)\delta\phi\nonumber\\
-\[u''(\xi)+\ymyo \normyzero w''(\xi)\]\frac{\dd \delta\phi}{\dd \xi}
-\[\normyzero v''(\xi)-\ymyo u'''(\xi)\]
{}&\frac{\dd \delta\phi}{\dd \normy}=
\nonumber\\
\left\{\[f(\xi)-\ymyo\normyzero g'(\xi)\]
\frac{\dd}{\dd \normy}+
\[\normyzero g(\xi)+\ymyo h(\xi)\]
\frac{\dd}{\dd \xi}\right\}
\(\frac{\dd^2}{\dd \xi^2}+
\aspratsq \frac{\dd^2}{\dd\normy^2}\)\delta\psi\nonumber\\
-\[\normyzero g''(\xi)+\ymyo h''(\xi)\]
\frac{\dd \delta\psi}{\dd\xi}
-\[f''(\xi)+\ymyo\normyzero g''\]
\frac{\dd \delta\psi}{\dd\normy}+{}&\mathcal O\(\ymyo^2\),
\end{align}
\end{widetext}
where we have used the normalizations defined in \eqs{normalizations}{time_lengths_norm}.

In the case of $y_0=0$, the above equations include the following effects which 
were absent in Paper I: 
\begin{itemize}
\item in \eq{lin_ohm}, the term proportional to 
  $h(\xi)$, represents the effect of the 
  reconnected magnetic field;
\item in \eq{lin_vort}, the terms proportional to 
  $u''(\xi)$ and $u'''(\xi)$ on the left-hand side, which represent the vorticity 
  of the equilibrium flow; 
\item the term proportional to $h''(\xi)$ on the 
  right-hand side of \eq{lin_vort}, which is the contribution to the equilibrium-current 
  gradient from the reconnected magnetic field.
\end{itemize}

Further progress at this point is hindered by the fact that these 
equations contain explicit dependences on the $y$ variable and cannot, 
therefore, be Fourier transformed in this direction. 
To address this difficulty, let us compare the magnitudes of the first and 
third terms on the left-hand sides of these equations:
\be
\frac{\dd/\dd \tau}{\ymyo~\dd/\dd \normy}
\sim \frac{\gamma}{\ymyo \kappa},
\ee
where $\gamma$  is the
growth rate of the anticipated instability at $\normy=\normyzero$ normalized to the 
Alfv\'enic shearing rate $\Gamma_0$ and
\be
\kappa=k\Lsheet
\ee
is the normalized wavenumber of the 
perturbation at that location.
Thus, the third term can be ignored if the analysis is restricted to patches of 
the current sheet whose extent in the $y$-direction is such that
\be
\label{deltay}
\ymyo \ll \gamma/\kappa.
\ee
This approach is valid provided that 
\be
\label{kappa_deltay}
\kappa \ymyo\gg1.
\ee
In other words, the domain in the $y$-direction is divided into smaller patches, and 
the linear analysis performed locally in of each of these. A WKB approach 
remains valid provided that asymptotically many wave lengths fit in each of these 
patches. \Eqs{deltay}{kappa_deltay} imply that we seek solutions such that
\be
\label{order_gamma}
\frac{\dd}{\dd \tau}\sim\gmax\gg 1,\qquad
\frac{\dd}{\dd \normy}\sim \kmax\gg 1.
\ee
These are  \textit{a priori} assumptions, which will be later justified 
by our ability to obtain such solutions.

Under these approximations,~\eqs{lin_ohm}{lin_vort} become
\begin{widetext}
\begin{align}
\label{lin_ohm_simple}
\frac{\dd \delta\psi}{\dd \tau}+
\normyzero v(\xi)\frac{\dd \delta \psi}{\dd \normy}
+u(\xi)\frac{\dd \delta \psi}{\dd \xi}
%
-\normyzero g(\xi)\frac{\dd \delta\phi}{\dd \xi}
-f(\xi)\frac{\dd \delta \phi}{\dd \normy}=
\(\frac{\dd^2}{\dd \xi^2}+\aspratsq\frac{\dd^2}{\dd\normy^2}\)\delta\psi,
\end{align}
\begin{align}
\label{lin_vort_simple}
\left\{\frac{\dd}{\dd \tau}+
\normyzero v(\xi)\frac{\dd}{\dd\normy}+
u(\xi)\frac{\dd}{\dd\xi}\right\}
\(\frac{\dd^2}{\dd \xi^2}+
\aspratsq \frac{\dd^2}{\dd\normy^2}\)&\delta\phi  
-u''(\xi)\frac{\dd \delta\phi}{\dd \xi}
-\normyzero v''(\xi)\frac{\dd \delta\phi}{\dd \normy}=
\nonumber\\
\[f(\xi)\frac{\dd}{\dd \normy}+\right.
{}&\left.\normyzero g(\xi)\frac{\dd}{\dd \xi}\]
\(\frac{\dd^2}{\dd \xi^2}+
\aspratsq \frac{\dd^2}{\dd\normy^2}\)\delta\psi  
-\normyzero g''(\xi)\frac{\dd \delta\psi}{\dd\xi}
-f''(\xi)\frac{\dd \delta\psi}{\dd\normy}.
\end{align}
\end{widetext}
Note that the functions $h(\xi)$ and $w(\xi)$ have dropped out.

One can now look for linear 
modes of the form $\exp(-i\omega \tau)$. 
In the midplane of the current sheet (i.e., $\xi=0$), the plasma 
is flowing outwards at some fraction of the Alfv\'en speed, 
$\normyzero v_0$, where $v_0=v(\xi=0,\normyzero)$.
Let us take this into consideration explicitly and set
\be
\label{moving frame transf}
\omega=\kappa \normyzero v_0+i\gamma.
\ee
We therefore look for solutions to Eqs.~(\ref{lin_ohm_simple}--\ref{lin_vort_simple}) in the
form:
\bea
\delta\psi&=&\Psi(\xi)e^{\[\gamma-
i\kappa\normyzero v_0\]\tau+i\kappa\normy},\\
\delta\phi&=&-i\Phi(\xi)e^{\[\gamma-
i\kappa\normyzero v_0\]\tau+i\kappa\normy}.
\eea
We are ignoring the time dependence of $\kappa$ due to 
the background flows, because  this variation will occur on a much longer 
timescale than that of the expected growth rate of the instability
\footnote{The time dependence of $\kappa$ induced by the background shear flow 
becomes important at low values of $S$ and has been measured numerically in 
Ref.~[\onlinecite{ni_linear_2010}].}.
Introducing the parameter 
\be
\lambda=\gamma/\kappa,
\ee
we obtain
\be
\label{linear_ohm}
\begin{split}
\Bigr\{\lambda-{}&
i\normyzero\[v_0-v(\xi)\]\Bigl\}\Psi
+\frac{u(\xi)}{\kappa}\Psi'\\
&+i\frac{\normyzero}{\kappa}g(\xi)\Phi'
-f(\xi)\Phi=
\frac{1}{\kappa}\(\Psi''-\kappa^2\epsilon^2\Psi\),
\end{split}
\ee
\be
\label{linear_momentum}
\begin{split}
\Bigr\{\lambda{}&-
i\normyzero\[v_0-v(\xi)\]\Bigl\}
\(\Phi''-\kappa^2\epsilon^2\Phi\)\\
&+\frac{u(\xi)}{\kappa}\(\Phi'''-\kappa^2\epsilon^2\Phi'\)
-\frac{u''(\xi)}{\kappa}\Phi'
-i\normyzero v''(\xi)\Phi=\\
&\quad-f(\xi)\(\Psi''-\kappa^2\epsilon^2\Psi\)
+i\frac{\normyzero}{\kappa} g(\xi)
(\Psi'''-\kappa^2\epsilon^2\Psi')\\
&\quad\quad-i\frac{\normyzero}{\kappa}g''(\xi)\Psi'
+f''(\xi)\Psi.
\end{split}
\ee
This is the set of equations that will be solved in \secsref{sec:lin_theory}{KH}.
\section{Plasmoid instability}
\label{sec:lin_theory}

We will now proceed to solve \eqs{linear_ohm}{linear_momentum} in three different regions: 
the ``external'' (global) region where $\deltacs\ll x\ll L$ (i.e., $1\ll\xi\ll S^{1/2}$), 
the ``outer'' region (the SP current sheet), 
where $x\sim\deltacs$ ($\xi\sim1$) and 
finally  the ``inner'' region (the inner layer inside the current sheet where plasmoids form), 
where $x\ll\deltacs$ ($\xi\ll1$). 
Outside the current sheet (i.e., in the 
external region), the plasma is ideal; inside the current sheet (i.e., 
in the outer and inner regions), resistive effects cannot be neglected.
In addition to the assumptions of \eq{order_gamma}, we will 
require here that $\lambda=\gamma/\kappa\ll 1$. This ordering is indeed 
satisfied by the fastest growing mode, as we will confirm {\it a posteriori}.

\subsection{External Region: $|\xi|\gg 1$}
\label{ext_reg}
This is the upstream region outside the current layer, i.e., 
$x\gg \deltacs$.
Here, we expect the equilibrium profiles to behave as:
\begin{align}
v(\xi)&\rightarrow 0, & u(\xi)&\approx \mp\uinf,\\
\label{f_g_ext}
~f(\xi)&\approx\pm \finf, &
g(\xi)&\approx\pm g_\infty'\xi,
\end{align}
where $\uinf,~\finf,~g_{\infty}'$ are functions of $\normyzero$ only, taken to be 
of order unity (see Appendix~\ref{equilibrium}); of these, we will 
discover that only $\finf$ matters for the calculation of the instability.
In the above expressions, the upper sign applies to $\xi>0$, 
and the lower sign to $\xi<0$ (so as to observe the expected parities 
of the equilibrium, namely that $u(\xi)$ and $f(\xi)$ are odd in $\xi$, 
and $g(\xi)$ is even).
The linear dependence of $g(\xi)$ on $\xi$ for large $\xi$ might not 
be obvious at first glance and is derived in Appendix~\ref{equilibrium}.

We will make the \textit{a priori} assumption that the terms 
proportional to $u(\xi)$ or $u''(\xi)$ 
and to $g(\xi)$ or $g''(\xi)$ in \eqs{linear_ohm}{linear_momentum} are negligible in 
this region, and then show that this is indeed the case. In the absence 
of these terms, we obtain:
\bea
\label{phi_ext}
\(\lambda-i\normyzero v_0\)\Psi&=&\pm\finf\Phi,\\
\(\lambda-i\normyzero v_0\)
\(\Phi''-\kappa^2\epsilon^2\Phi\)&=&\mp\finf(\Psi''-\kappa^2\epsilon^2\Psi),
\eea
which can be easily combined to yield:
\be
\[\finf^2-(\lambda - i \normyzero v_0)^2\]\(\Psi''-\kappa^2\epsilon^2\Psi\)=0.
\ee
The general solution to this equation is simply
\be
\label{psi_ext}
\Psi^{\pm}=C_3^{\pm}e^{\mp\kappa\epsilon\xi},
\ee
where $C_3^{\pm}$ are integration constants and $\pm$ refers to $\xi\gtrless 0$
[in unscaled units:
$\Psi^{\pm}=C_3^\pm \exp(\mp kx)$].

We can now check the assumption about the smallness of the terms 
proportional to $u(\xi),~u''(\xi),~g(\xi)$ and $g''(\xi)$.
From \eq{psi_ext} and \eq{phi_ext} we see that, for arbitrary $y_0$, we have 
$\Psi'/\Psi \sim \Phi'/\Phi \sim \kappa \epsilon$.
Let us then compare the magnitudes of the third and first
terms on the left-hand side of \eq{linear_ohm} [the same reasoning applies to 
\eq{linear_momentum}]:
\be
\frac{(\uinf/\kappa) \Psi'}{\lambda \Psi}\sim \lambda^{-1} \epsilon,
\ee
which is small provided that $\lambda \gg \epsilon$, a condition we will later see 
is satisfied by the fastest growing mode. 
With respect to terms involving $g(\xi)$, the ratio of the magnitudes of the
fourth and fifth terms on the left-hand side of \eq{linear_ohm} [and similarly for 
the second and first terms on the right-hand side of~\eq{linear_momentum}]
is
\be
\normyzero \frac{\xi}{\finf^2\kappa}\frac{\Phi'}{\Phi}
\sim \xi \epsilon
\ee
This is  again small provided that $\xi \ll \epsilon^{-1} \sim S^{-1/2}$;
we shall find that the fastest growing wave number $\kmax\sim\epsilon^{-3/4}$ and thus 
the eigenfunction, \eq{psi_ext}, decays before the condition $\xi\epsilon\ll1$ breaks down.
Note also that the expression for $g(\xi)$ given in \eq{f_g_ext} is not 
expected to hold for $\xi>\epsilon^{-1}$ (i.e., $x>\Lsheet$): the reconnected magnetic 
field does not grow unbounded as $\xi\rightarrow\infty$). 

\subsection{Outer Region: $|\xi|\sim 1$}
\label{sec_outer_region}
This region represents the SP current sheet itself, i.e., $x\sim\deltacs$.
Here, the functions $u(\xi),~v(\xi),~f(\xi)$ and $g(\xi)$ are simply assumed 
to be $\sim \mathcal O(1)$.
We find that terms proportional to $u(\xi)$ and $g(\xi)$ or to their derivatives
are again negligible. For example, consideration of the same terms as in the previous 
section leads to:
\bea
\frac{u\Psi'/\kappa}{\lambda\Psi}&\sim&\gamma^{-1}\ll1\\
\frac{\normyzero g\Phi'/\kappa}{f\Phi}&\sim& 
\normyzero\kappa^{-1}\ll1,
\eea
and similarly for the others. Therefore, to lowest order in $\epsilon$, 
\eqs{linear_ohm}{linear_momentum} become
\be
\left\{\lambda-i\normyzero \[v_0-v(\xi)\]\right\}\Psi=
f(\xi)\Phi,
\ee
\be
\begin{split}
\{\lambda-i\normyzero &\[v_0- v(\xi)\]\}\(\Phi''-\kappa^2\epsilon^2\Phi\)
-i\normyzero v''(\xi)\Phi=\\
&-f(\xi)\Psi''+\[f''(\xi)+
\kappa^2\epsilon^2 f(\xi)\]\Psi.
\end{split}
\ee
Combining these equations results in the following eigenvalue problem:

\begin{widetext}
\be
\label{eigen_B}
\Psi''-\[\frac{f''(\xi)}{f(\xi)}+\kappa^2\epsilon^2\]\Psi=
-\frac{\lambda-i\normyzero\[v_0-v(\xi)\]}{f(\xi)}
\left\{\(\frac{d^2}{d\xi^2}-\kappa^2\epsilon^2\)
\frac{\lambda-i\normyzero\[v_0-v(\xi)\]}{f(\xi)}\Psi-
\frac{i\normyzero v''(\xi)}{f(\xi)}\Psi\right\},
\ee
\end{widetext}
subject to boundary conditions given by the external solution, \eq{psi_ext}, and the 
requirement (for the plasmoid instability) that $\Psi$ be an even function.
Writing the  solution in the form $\Psi(\xi) = f(\xi) \chi(\xi)$, \eq{eigen_B} becomes:
\be
\label{chi_form}
\frac{d}{d\xi}\[V(\xi)\chi'(\xi)\]=\epsilon^2\kappa^2V(\xi)\chi(\xi),
\ee
where the ``potential'' $V(\xi)$ is
\be
\label{V_pot}
V(\xi)=f^2(\xi)+\[\lambda-iy_0(v_0-v(\xi))\]^2.
\ee

Note that in the case $y_0\ll 1$ all terms on the 
right-hand side of \eq{eigen_B} are small compared to the first two terms on the left-hand side. 
In that case,
we recover to the problem solved in Paper I [except here $f(\xi)$ remains unspecified].
For the general case $\normyzero\sim 1$, an exact solution can 
be obtained provided that the terms proportional to $\kappa^2\epsilon^2$ can be 
neglected (we will later check the validity of this assumption; 
for now, let us call attention to what it means:
$\kappa\epsilon = k\deltacs$, so the assumption $\kappa\epsilon\ll 1$ implies 
that the wavelength of the expected instability is much longer than the 
current sheet thickness). 
Thus, we  neglect the right-hand side of \eq{chi_form} and find the solution
\be
\label{sol_regII_general}
\Psi^\pm(\xi)=C_1^\pm f(\xi)+
C_2^\pm f(\xi)
\int_{\xi_0}^\xi\frac{d\xi'}{V(\xi')}.
\ee
where $\pm$ refers to $\xi\gtrless 0$, $C_1^\pm,~C_2^\pm$ are constants in integration 
and $\xi_0$ is an arbitrary number of order unity (different choices of 
$\xi_0$ will produce subdominant corrections to $C_1^\pm$).

For the plasmoid instability, we expect $\lambda \ll 1$, so
this solution simplifies to
\be
\begin{split}
\label{sol_regII}
\Psi^\pm=C_1^\pm f(\xi)+\\
C_2^\pm f(\xi)&
\int_{\xi_0}^\xi\frac{d\xi'}{f^2(\xi')-
\normyzero^2\[v_0-v(\xi')\]^2}.
\end{split}
\ee

We now match this expression to the external solution, \eq{psi_ext}, in 
the region
$1 \ll \xi \ll (\kappa \epsilon)^{-1}$, or, equivalently, in dimensional form,
$\delta_{CS} \ll x \ll 1/k$.
In this region,
$v(\pm\xi)\ll v_0$ and $f'(\pm\xi)\ll1$, implying that
$|f(\pm\xi)|\approx\finf$ and the integral is dominated by the upper limit.
We obtain
\bea
C_3^\pm&=& -\frac{C_2^\pm \finf}{\fsqinf-\normyzero^2 v_0^2} 
\frac{1}
{\kappa\epsilon},\\
\label{eq:c_3}
C_1^\pm&=&\mp
\frac{C_2^\pm}{\fsqinf-\normyzero^2v_0^2}
\frac{1}{\kappa\epsilon}.
\eea

For $\xi\ll1$, $f(\xi)\approx f_0'\xi$ 
and $v(\xi)=v_0-v''_0\xi^2/2$. Thus, to lowest order in $\xi$, the integrand 
in \eq{sol_regII} becomes 
$1/({f'_0}^2\xi^2)$.
The integral is therefore again dominated by the upper limit and 
we obtain:
\be
\Psi^\pm(0)=-\frac{C_2^\pm }{f_0'}.
\ee
Demanding that $\Psi$ be an even function of $\xi$
 yields
\be
C_2^+=C_2^-=-{f'_0}\Psi(0).
\ee
Thus, the outer region solution is:
\be
\label{out_psi}
\begin{split}
\Psi^\pm(\xi)=f'_0&\Psi(0)f(\xi)
\left\{\pm\frac{1}{\fsqinf-\normyzero^2v_0^2}
\frac{1}{\kappa\epsilon}\right.\\
&\left.-\int_{\xi_0}^\xi \frac{d\xi'}{f^2(\xi')-
\normyzero^2\[v_0-v(\xi')\]^2}\right\}.
\end{split}
\ee
Note that the second term of the above expression ensures that the eigenfunction 
remains finite at $\xi=0$. 

As usual in tearing-mode-type calculations, let us now introduce 
the standard instability parameter
\be
\DD=\frac{\Psi'(+0)-\Psi'(-0)}{\Psi(0)}.
\ee
Using \eq{out_psi}, we obtain~\footnote{Note that in the limit $\normyzero=0$, we recover 
Eq.~(15) of Paper I, with the exceptions that (i) here $f'_0$ refers to a generic equilibrium and
(ii) since the profiles are now continuous and we are no longer matching at a point, the 
subdominant contribution to $\DD$ present in Paper I no longer appears in \eq{delta_prime_simple}.}
\be
\begin{split}
\label{delta_prime_simple}
\DD={}&\frac{2}{\kappa\epsilon}
\frac{{f'_0}^2}{\fsqinf-\normyzero^2v_0^2}.
\end{split}
\ee
We stress that the functional dependence of $\DD$ on
$\normyzero$ is both explicit and implicit, as $f'_0,~\finf$ and $v_0$ 
are all functions of $\normyzero$.

\subsection{Inner Region: $|\xi|\ll1$}
\label{inner_region}
This region is the internal layer inside the SP current sheet, i.e., $x\ll\deltacs$.
We begin by noting that, again, independently of the specific functional form of the SP 
equilibrium, the symmetries of the problem 
are such that, for $|\xi|\ll1$, 
the equilibrium profiles can be approximated as
\bea
f(\xi)&=&f'_0\xi+\mathcal O(\xi^3),\quad
g(\xi)=g_0+\mathcal O(\xi^2),\\
u(\xi)&=&u'_0\xi+\mathcal O(\xi^3),\quad
v(\xi)=v_0+\mathcal O(\xi^2),
\eea
where $f'_0,~u'_0,~g_0,~v_0$ are constants with respect to $\xi$ but depend on $\normyzero$.
In this region, the relative magnitudes of the different terms in 
\eqs{linear_ohm}{linear_momentum} can be reduced to one of the following cases:
\bea
\frac{u\Psi'}{\kappa}\frac{1}{\lambda \Psi}&\sim& 
\frac{u'_0 \deltain \Psi}{\kappa \deltain}\frac{\kappa}{\gamma \Psi}\sim \frac{1}{\gamma},\\
\frac{\normyzero g \Phi'}{\kappa}\frac{1}{f\Phi}&\sim& \frac{g_0}{\kappa f'_0\deltain^2}
\sim\frac{1}{\kappa\deltain^2},\\
\label{keep}
\frac{\Psi''}{\kappa}\frac{1}{\lambda\Psi}&\sim&\frac{1}{\gamma\deltain^2},\\
\frac{u''\Phi'}{\kappa}\frac{1}{\lambda\Phi''}&\sim&\frac{\deltain^2}{\gamma},\\
\frac{\normyzero v'' \Phi}{\lambda\Phi''}&\sim&\frac{\kappa\deltain^2}{\gamma},\\
\frac{\normyzero g'' \Psi'}{\kappa}\frac{1}{ f \Psi''}&\sim& \frac{1}{\kappa}.
\eea
Except for \eq{keep}, all these ratios can be shown {\it a posteriori} to be small.
Thus, to lowest order,
\eqs{linear_ohm}{linear_momentum} become
\bea
\label{in_reg_Psi}
\lambda\Psi-f'_0\xi\Phi&=&\frac{1}{\kappa}\Psi'',\\
\label{in_reg_Phi}
\lambda\Phi''&=&-f'_0\xi\Psi''.
\eea

These equations are mathematically the same as the equations for  the tearing mode in the inner region,
except that here the role of the small parameter is played by $1/\kappa$ rather than resistivity.
Since $\DD\deltain$ is not expected to be small,
the constant-$\Psi$ approximation~\cite{furth_finite-resistivity_1963} cannot be used. 
Instead, this eigenvalue problem is mathematically equivalent to the one 
solved by Coppi \textit{et al.}~for the resistive internal kink mode~\cite{coppi_1976}.
The resulting dispersion relation is
\be
\begin{split}
\label{disprel}
-\frac{\pi}{8}\(\kappa f'_0\)^{1/3}\Lambda^{5/4}
\frac{\Gamma\[\(\Lambda^{3/2}-1\)/4\]}
{\Gamma\[\(\Lambda^{3/2}+5\)/4\]}=\DD=\\
\frac{2}{\kappa\epsilon}
\frac{{f'_0}^2}{\fsqinf-\normyzero^2v_0^2},
\end{split}
\ee
where $\Gamma$ is the gamma function and
$\Lambda=\gamma {f'_0}^{-2/3}\kappa^{-2/3}$. 
The width of the inner region is
\be
\label{delta_in_def}
\deltain=\gamma^{1/4}\kappa^{-1/2}{f'_0}^{-1/2},
\ee
which can be used to confirm the smallness of the terms neglected in 
deriving \eqs{in_reg_Psi}{in_reg_Phi}.
\subsection{Solution of the Dispersion Relation, \eq{disprel}}
\label{sec:analysis_disprel}
The dispersion relation,~\eq{disprel}, has two relevant limits.
For  $\Lambda\ll1$,
\begin{align}
\label{Lambda_small}
\gamma& \approx\[-\frac{16}{\pi}\frac{\Gamma (5/4)}{\Gamma (-1/4)}\]^{4/5}
\frac{{f'_0}^2}{
\(\finf^2-\normyzero^2v_0^2\)^{4/5}}
~\kappa^{-2/5}\epsilon^{-4/5},\nonumber\\
&\approx 0.95 \frac{{f'_0}^2}{
\(\finf^2-\normyzero^2v_0^2\)^{4/5}}
~\kappa^{-2/5}\epsilon^{-4/5}.
\end{align}
This is valid provided that $\kappa\gg\epsilon^{-3/4}$ (but also $\kappa\epsilon\ll1$, 
as required by our earlier assumptions).

On the other hand, taking $\Lambda\rightarrow1-$ (from below), we obtain
\be
\label{Lambda_1}
\gamma= (f'_0 \kappa)^{2/3}-
\frac{\sqrt{\pi}}{3}
\frac{\fsqinf-\normyzero^2v_0^2}{f'_0}
\kappa^2\epsilon,
\ee
valid for $\kappa\lesssim\epsilon^{-3/4}$.
The scaling of the fastest growing wave number can be determined
by balancing the two terms on the right-hand side of the 
above expression. The exact result can be obtained by solving the 
equation $d\gamma/d\kappa=0$ in the limit $\Lambda\rightarrow 1-$. 
We obtain
\bea
\label{k_max}
\kmax&=&
\(\frac{1}{\sqrt{\pi}}\frac{{f'_0}^{5/3}}
{\fsqinf-\normyzero^2v_0^2}\)^{3/4}\epsilon^{-3/4},\\
\label{gamma_max}
\gmax&=&
\frac{2}{3\pi^{1/4}} \sqrt{\frac{{f'_0}^3}{\fsqinf-\normyzero^2v_0^2}}
\epsilon^{-1/2}.
\eea 
 
The  scalings with $\epsilon$ are the same as those 
derived in Paper I, although here they have been obtained for a general SP equilibrium.
In other words, this linear theory predicts that current sheets are 
unstable to a super-Alfv\'enic instability whose growth rate increases with 
the Lundquist number, $\gmax \sim S^{1/4}$ 
[recall that $\epsilon = S^{-1/2}$; see \eq{epsilon_def}].
A plasmoid chain forms inside a region of width 
$\deltain\sim S^{-1/8}\deltacs$, with the number of plasmoids scaling 
as $\kmax\sim S^{3/8}$. These scalings justify the ordering assumptions 
employed in deriving these results, i.e., $\gmax\gg 1$, $\kmax \gg 1$ and
$\lambda_{\rm max}\sim \gmax/\kmax\sim\epsilon^{1/4}$, so, $\epsilon \ll \lambda_{\rm max} \ll 1$.

Let us now analyze the dependence of $\gmax$ and $\kmax$ on the position $\normyzero$ along the sheet.
Note first that the instability vanishes at the locations where 
the equilibrium current $f'_0=0$ --- the end points of the SP current sheet.
Note also that since we have dropped corrections of order $\epsilon^{1/2}$ in our 
derivation [for example, the terms proportional to 
$\kappa^2\epsilon^2$ in \eq{chi_form}],  the terms proportional to $\normyzero^2$ 
in the above expressions 
are only to be kept if $\normyzero\gg\epsilon^{1/4}$. 
For values of $\normyzero\lesssim\epsilon^{1/4}$, all $\normyzero$ corrections are negligible
to lowest order and  \eqs{k_max}{gamma_max} simplify to
yield
\bea
\label{kmax_zero}
\kmax&=&\pi^{-3/8}~\bar E_0^{5/4}\epsilon^{-3/4},\\
\label{gamma_max_zero}
\gmax&=&\frac{2}{3\pi^{1/4}}~\bar E_0^{3/2}\epsilon^{-1/2},
\eea
where $\bar E_0=\Lsheet E_0/(B_0^2\deltacs)$  is the normalized background electric 
field, and we have used the relationship $f'_0 = \bar E_0$,
which follows from \eq{equilib_fprime0} in the limit $\normyzero\lesssim \epsilon^{1/4}$ [note 
that $\finf(\normyzero=0)=1$].

In the opposite case of $\normyzero\gg\epsilon^{1/4}$,  
the dependence of $\kmax$ and $\gmax$ on $\normyzero$ is a
nontrivial function of the specific values of the equilibrium coefficients,
which are all functions of $\normyzero$
(this does not affect the scaling of 
$\kmax$ and $\gmax$ with $\epsilon$). 
For $\normyzero\ll 1$, 
exact values of the coeficients of the 
Taylor expansion of the equilibrium around $\xi=0$ were derived 
semi-analytically by Uzdensky and Kulsrud~\cite{uzdensky_viscous_1998} assuming 
a Syrovatskii-like upstream magnetic field [\eq{KH_syro}] --- see Appendix~\ref{equilibrium}.
Using those coefficients and the relationship 
$f'_0=\bar E_0-\normyzero^2v_0g_0$ 
derived in Appendix~\ref{equilibrium}, a Taylor expansion
of \Eqs{k_max}{gamma_max} in $\normyzero$ yields:
\bea
\label{k_max_uz}
\kmax(|\normyzero|\ll1)&\approx&
\(0.56+0.10\normyzero^2\)\epsilon^{-3/4},\\
\label{gamma_max_uz}
\gmax(|\normyzero|\ll1)&\approx&
\(0.71+0.73\normyzero^2\)\epsilon^{-1/2}.
\eea

These results reveal a perhaps unexpected feature of the plasmoid instability: 
 both $\kmax$ and $\gmax$ {\it increase} with distance from the center 
of the sheet. The same conclusion is easily deduced for arbitrary 
$\normyzero\sim 1$: even though in that case one is forced to retain the 
full expressions for $\kmax$ and $\gmax$, \eqs{k_max}{gamma_max}, it is also true that, 
under very general conditions, one expects $\finf$ to be a 
decreasing function of $\normyzero$ (for example, 
a standard Syrovatskii ideal current sheet solution~\cite{syrovatskii_formation_1971} 
yields $\finf = \sqrt{1-\normyzero^2}$; see Appendix~\ref{equilibrium}).

A problem thus arises:
it is possible that a location $\normyzero=\normyzerocrit\sim \mathcal O(1)$
exists inside the current sheet where 
$\finf^2-\normyzerocrit^2v_0^2=0$, and our solution breaks down 
($\DD, \gamma, \kappa \rightarrow +\infty$).
It is clear that while approaching that point, $\kmax$ will 
get to be so large that terms of order $\kmax\epsilon$, or $\kmax^2\epsilon^2$,
can no longer be neglected and thus our ordering assumptions become invalid.
For values of $|\normyzero|>\normyzerocrit$, our solution is again physical,  
but the second term on the right-hand side of~\eq{Lambda_1}
equation changes sign, implying that $\gamma$ grows with $\kappa$
and the value of the fastest-growing wave number cannot be deduced from this equation.
This means that terms proportional to $\kappa^2\epsilon^2$ are still
necessary to determine the fastest growing mode.
Physically, the fact that the condition $\kmax\epsilon\sim 1$ 
is met somewhere in the sheet means that 
at those locations the wavelength of the perturbation
becomes comparable to the current sheet thickess, whereas for 
$|\normyzero| \ll \normyzerocrit$ it was much longer.

It is easy to understand why the location $\finf^2-\normyzerocrit^2v_0^2=0$ should 
be special: this is where 
the midplane outflow speed (i.e., $u_y$ measured at $x=0$) matches the Alfv\'en velocity associated 
with the upstream 
magnetic field (measured at $x\sim\deltacs$), i.e., it is
the Alfv\'en Mach point of 
the system. The background outflow velocity profile is strongly sheared (in the $x$-direction) 
inside the current sheet.
 Such a profile would be unstable to the Kelvin-Helmholtz (KH)
instability, were it not for the stabilizing effect provided by the 
(flow-aligned) background magnetic field $B_y$~\cite{chandrasekhar_hydrodynamic_1961}.
However, whereas the flow grows in magnitude with increasing $y$, $B_y$ does the 
opposite (see the discussion of \secref{heur_KH}). 
The Alfv\'en Mach point is where the two amplitudes match. Beyond 
that point, the magnetic field is no longer sufficiently strong to provide stability, and 
we should thus expect the plasmoid instability to morph into the KH instability.
The increase of $\gmax$ and $\kmax$ along the sheet is thus a reflection of the 
fact that the current sheet is increasingly unstable to the KH mode.  
For $\normyzero\gtrsim \normyzerocrit$ the plasmoid instability is replaced by the 
KH instability as the most unstable mode.
The most unstable wave number also 
grows along the sheet because, as is well known~\cite{chandrasekhar_hydrodynamic_1961}, 
the growth rate of the 
KH instability peaks at $\kappa \epsilon\sim1$ (i.e., $k\deltacs\sim1$, where 
$\deltacs$ is the scale of the cross-sheet velocity shear).

To summarize, the analytical dispersion relation,~\eq{disprel},
accurately describes the plasmoid instability of the 
current sheet for  $-\normyzerocrit<\normyzero<\normyzerocrit$. 
For values of $\normyzero$ outside this interval, it becomes necessary to keep
terms proportional to $\kappa^2\epsilon^2$ in order to calculate the fastest 
growing mode, which is no longer the plasmoid instability, but the KH instability.
Retaining the $\kappa^2\epsilon^2$  terms analytically is difficult; however, 
we have been able to 
address the full problem by a direct numerical solution of the linearised equations. 
Results are 
shown in~\secref{numerics}.
Before discussing those, though, let us present an analytical derivation of the 
KH instability of a current sheet valid in the long-wavelength limit $\kappa\epsilon\ll 1$.

\section{Kelvin-Helmholtz instability of the layer}
\label{KH}
In this section, we derive an analytical dispersion relation of the 
Kelvin-Helmholtz (KH) instability of the current sheet valid for perturbations 
whose wave number is $\kappa\ll\epsilon^{-1}$.
This ordering of $\kappa$ is not expected to capture the fastest growing mode, as suggested by the 
heuristic derivation of \secref{heur_KH} ($\kmax\sim\epsilon^{-1}$), but we are unable to 
obtain an analytic solution valid for $\kappa\epsilon\sim 1$. 
The derivation presented here, however, does reveal a number of interesting 
features of the KH instability of the current sheet. 
The direct numerical solution presented in the next section does of course cover 
all values of $\kappa$.

We first remind the reader that in a SP reconnection configuration, the outflow velocity profile is 
maximum at the midplane ($x=0$) of the sheet and decays to zero away from it. 
There are, therefore, 
two shear layers on each side of the sheet, at $x\sim \pm\deltacs$, where the KH instability 
may develop --- see \fig{fig:KH_cartoon}. 
These two shear layers will push magnetic fields of opposite sign on each side of $x=0$ 
towards each other. This  will create a current sheet at $x=0$.
Thus, resistivity can play an important role in this mode, 
not at the KH layers themselves, $x\simeq\pm\deltacs$, but at $x=0$, where the shear layers 
interact.
This situation is then conceptually similar to the forced reconnection problem treated by 
Hahm \& Kuslrud~\cite{hahm_forced_1985} (the Taylor problem), 
where perturbations at some distant boundaries on each side of a 
rational surface induce reconnection at that surface.
Here, the KH instability can be thought of as 
the equivalent of those perturbations at the boundaries, forcing reconnection at $x=0$.

In \cite{hahm_forced_1985} (see also~\cite{cole_forced_2004}) it was 
found that if the change in the boundaries occurs on a timescale much 
faster than the resistive one, there will be no reconnection in the 
early (linear) stage of evolution; 
the magnetic field will pile up until the current gets sufficiently large for the resistive 
term to become important.
In a somewhat similar fashion, here, whether or not the KH instability induces reconnection 
depends on how large 
its growth rate is; for low values of $\kappa\epsilon$, when its growth rate is lower, 
the pile up of the magnetic field is prevented by reconnection, 
which can proceed at a rate comparable to the growth rate of the KH instability.
However, for larger values of $\kappa\epsilon$, the KH instability will be faster 
than the rate at which reconnection can occur and we expect to find an ideal 
(i.e., non-reconnecting) mode (though reconnection is expected to start in the nonlinear stage, 
which we do not address here).

Our formal analysis of this problem again considers three asymptotic regions: 
the inner region ($|\xi|\ll 1$), 
the outer region (or flow shear layer, $|\xi|\sim 1$),
and the external region ($|\xi|\gg 1$). 
Since all the same orderings that led to the simplification of \eqs{linear_ohm}{linear_momentum}
in the previous section are still expected to hold here, the equations to solve in each 
of these regions are the same. In particular, the solution in the  
the external and outer regions remains unchanged and is given by Eqs.~(\ref{psi_ext}) 
and (\ref{sol_regII}), respectively.

In the inner region, we must solve \eqs{in_reg_Psi}{in_reg_Phi}.
There are two cases of interest: when
the right-hand side of \eq{in_reg_Psi} is 
important, and when it is not.
The former case occurs at low values of $\kappa\epsilon$, whereupon we simply recover 
the dispersion relation~\eq{Lambda_1}. 
This is similar to the resistive-kink-mode solution of Coppi \etal~\cite{coppi_1976}:
as $\normyzero$ increases and eventually becomes such that $\normyzero v_0 > \finf$, $\DD$ 
[\eq{delta_prime_simple}]
transitions from positive to negative via infinity; mathematically, this is equivalent to 
the well known transition 
from the very unstable (large-$\DD$) tearing mode to the resistive kink mode~\cite{coppi_1976}, 
except here reconnection is driven by the KH instability, rather than by the kink mode.

For larger values of $\kappa\epsilon$ (though still requiring that $\kappa\epsilon\ll1$; 
we will later 
determine how large $\kappa$ has to be for the following to hold), we can 
ignore the right-hand side of~\eq{in_reg_Psi} and obtain
\be
\label{psi_phi_KH_inner}
\Psi=\frac{f'_0 \xi}{\lambda}\Phi.
\ee
So, \eq{in_reg_Phi} becomes
\be
\Phi'' = -\frac{{f'_0}^2}{\lambda^2}\xi \(2\Phi' + \xi \Phi''\),
\ee
to be solved subject to the boundary condition $\Phi(0)=0$~\footnote{In theory, one could also 
look for solutions subject to the boundary condition $\Phi'(0)=0$, which would yield 
$\Phi(\xi)=const.\equiv\Phi_0$, and $\Psi(\xi)=f'_0\xi\Phi_0/\lambda$. 
Such a solution, however, cannot be 
matched to the one obtained in the outer region, \eq{psi_ext}, unless $C_2^{\pm}=0$, which 
implies $\Psi(\xi)=0$ in the entire domain.}. 
The solution is:
\be
\label{phi_KH_inner}
\Phi^{\pm}(\xi)= \frac{\Phi'_0\lambda}{f'_0}\arctan\(\frac{f'_0}{\lambda}\xi\),
\ee
where $\Phi'_0\equiv \Phi'(0)$. The inner layer width is, therefore, 
\be
\label{deltain_KH}
\delta_{\rm inner}=\frac{\lambda}{f'_0}.
\ee

Using \eq{phi_KH_inner} to substitute for $\Phi$ in \eq{psi_phi_KH_inner}, we obtain
\be
\label{psi_KH_inner}
\Psi^{\pm}(\xi)=\Phi'_0\xi\arctan\(\frac{f'_0}{\lambda}\xi\).
\ee
As expected this is a non-reconnecting mode: 
$\Psi(0)=0$ --- indeed, this solution is mathematically equivalent to the 
ideal-kink-mode solution found by Rosenbluth~\etal~\cite{rosenbluth_nonlinear_1973};
physically, the difference here is that the drive is the KH instability.

The solution, \eq{psi_KH_inner}, is now matched to the solution in the outer region,~\eq{sol_regII}. 
For $\xi\gg 1$, \eq{psi_KH_inner} becomes:
\be
\Psi^{\pm}(\xi)= \Phi'_0\xi\(\pm \frac{\pi}{2}-\frac{\lambda}{f'_0\xi}\).
\ee
Therefore, we have,
\bea
C_1^\pm &=&\pm \frac{\pi}{2}\frac{\Phi'_0}{f'_0},\\
C_2^\pm &=& \lambda \Phi'_0.
\eea
Substituting these expressions for $C_1^\pm,~C_2^\pm$ in \eq{eq:c_3}, we obtain the final dispersion 
relation:
\be
\label{KH_DR_simple}
\gamma = \lambda\kappa=\frac{\pi}{2} \frac{\normyzero^2v_0^2-\finf^2}{f'_0}\kappa^2\epsilon.
\ee
This expression shows that an 
unstable mode exists when $|\normyzero v_0| > |\finf|$, i.e., above
the Alfv\'en Mach point of the system, $|\normyzero|>\normyzerocrit$.
To determine its region of validity, let us use \eq{KH_DR_simple} and \eq{deltain_KH} to 
compare the the right-hand side of \eq{in_reg_Psi}, which we neglect in this derivation, to the 
first term on the left-hand side of that equation.
We then find that this solution is valid for $\epsilon^{-3/4}\ll\kappa\ll\epsilon^{-1}$.

Finally, we estimate the value of $\kappa=\kappa_{\rm tr}$ above which \Eq{KH_DR_simple} 
yields faster growth than the resistive dispersion relation, \eq{Lambda_1}.
This corresponds to a transition from a reconnecting to a non-reconnecting mode. 
Comparing the two expressions we find
\be
\label{KH_kappa_tr}
\kappa_{\rm tr}=\(\frac{\pi}{2}-\frac{\sqrt{\pi}}{3}\)^{-3/4}\frac{{f'_0}^{5/4}}
{\(\normyzero^2 v_0^2-\finf^2\)^{3/4}}\epsilon^{-3/4}.
\ee
Wave numbers such that $\kappa>\kappa_{\rm tr}$ are ideal, non-reconnecting modes (i.e., $\Psi(0)=0$).
This implies, in particular, that the fastest growing mode, $\kmax\sim\epsilon^{-1}$, 
is an ideal mode, 
as will be confirmed by the full numerical solution presented in the next section.

\section{Numerical results}
\label{numerics}
In this section, we compare the heuristic and the analytical results of 
sections \ref{sec:heuristic},~\ref{sec:lin_theory} and~\ref{KH} 
with the direct numerical solution of the full set of linearized equations:
\begin{align}
\label{IV_psi}
&\(\frac{1}{\kappa}\frac{\dd}{\dd \tau} + i \normyzero v(\xi)\) \Psi 
+\frac{u(\xi)}{\kappa}\Psi'\nonumber\\
&\quad+i\frac{\normyzero}{\kappa}g(\xi)\Phi'
-f(\xi) \Phi = 
\frac{1}{\kappa}\(\Psi''-\kappa^2\epsilon^2\Psi\),
\end{align}
\begin{align}
\label{IV_vort}
&\(\frac{1}{\kappa}\frac{\dd}{\dd \tau} + i\normyzero v(\xi)\)
(\Phi''-\kappa^2\epsilon^2\Phi) \nonumber\\
&\quad+\frac{u(\xi)}{\kappa}\(\Phi'''-\kappa^2\epsilon^2\Phi'\)
-\frac{u''(\xi)}{\kappa}\Phi'
-i\normyzero v''(\xi)\Phi = \nonumber\\
&-f(\xi)(\Psi''-\kappa^2\epsilon^2\Psi) 
+i\frac{\normyzero}{\kappa} g(\xi)
(\Psi'''-\kappa^2\epsilon^2\Psi')\nonumber\\
&\quad-i\frac{\normyzero}{\kappa}g''(\xi)\Psi'
+f''(\xi) \Psi,
\end{align}
where $\xi=x/\deltacs$, $\kappa=k\Lsheet$ and $\epsilon=\deltacs/\Lsheet=S^{-1/2}$.
These equations follow straightforwardly from \eqs{lin_ohm_simple}{lin_vort_simple}
after Fourier decomposing in the $y$-direction: 
$\delta\psi = \Psi(\xi)\exp(i\kappa y)$,  $\delta\phi = -i\Phi(\xi)\exp(i\kappa y)$.
The functions $f(\xi),~g(\xi),~u(\xi)$ and $v(\xi)$ are the normalized SP-like background 
equilibrium profiles: the reconnecting and reconnected magnetic field components
and the inflow and outflow velocity profiles, respectively 
(see \secref{sec:norms} for the normalizations adopted for them).
Explicit expressions for these functions, 
parametrized by the position along the sheet $\normyzero=y_0/\Lsheet$, are derived in 
Appendix~\ref{equilibrium}. For the equilibrium adopted, 
the Alfv\'en Mach point of the current sheet, defined by 
$u_y(\xi=0,\normyzerocrit)=B_y(\xi=\infty,\normyzerocrit)$,
occurs at $\normyzerocrit\approx 0.61$. Above this point, 
the upstream magnetic field is no longer able to stabilize the KH mode.

\Eqs{IV_psi}{IV_vort} are solved in a domain of size $-L_x\le\xi\le L_x$ using a 
second-order-accurate predictor-corrector numerical scheme. 
The boundary conditions 
are $\Psi(-L_x,t)=\Psi(L_x,t)=\Phi(-L_x,t)=\Phi(L_x,t)=0$.
The size of the simulation domain $L_x$ depends on $\kappa$, with lower values 
of $\kappa$ requiring 
larger domains [this is due to the behavior of the eigenfunction in the 
external region, $\Psi\sim e^{-\kappa\epsilon\xi}$; see \eq{psi_ext}].
Convergence tests were performed to
ensure that both the domain size and resolution were appropriate.
\subsection{Plasmoid and KH instabilities}
\begin{figure*}
\includegraphics[width=0.48\textwidth]{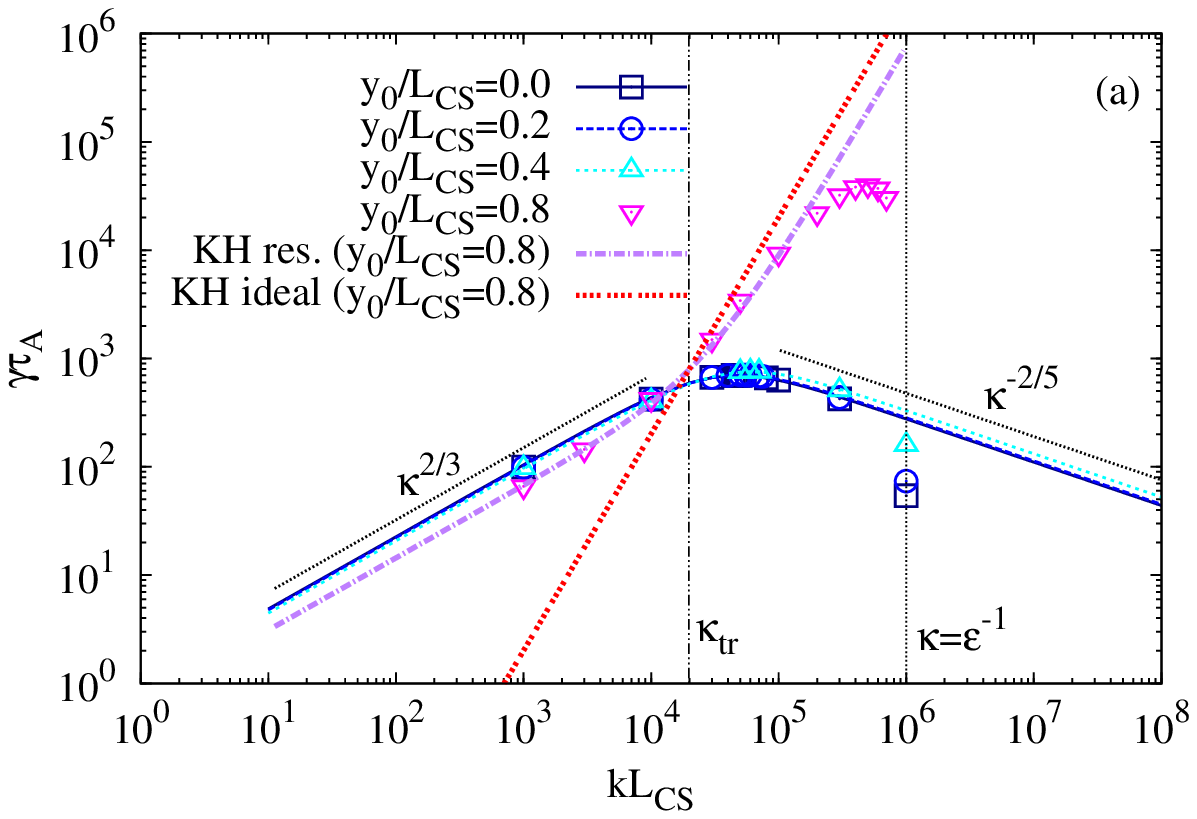}
\includegraphics[width=0.48\textwidth]{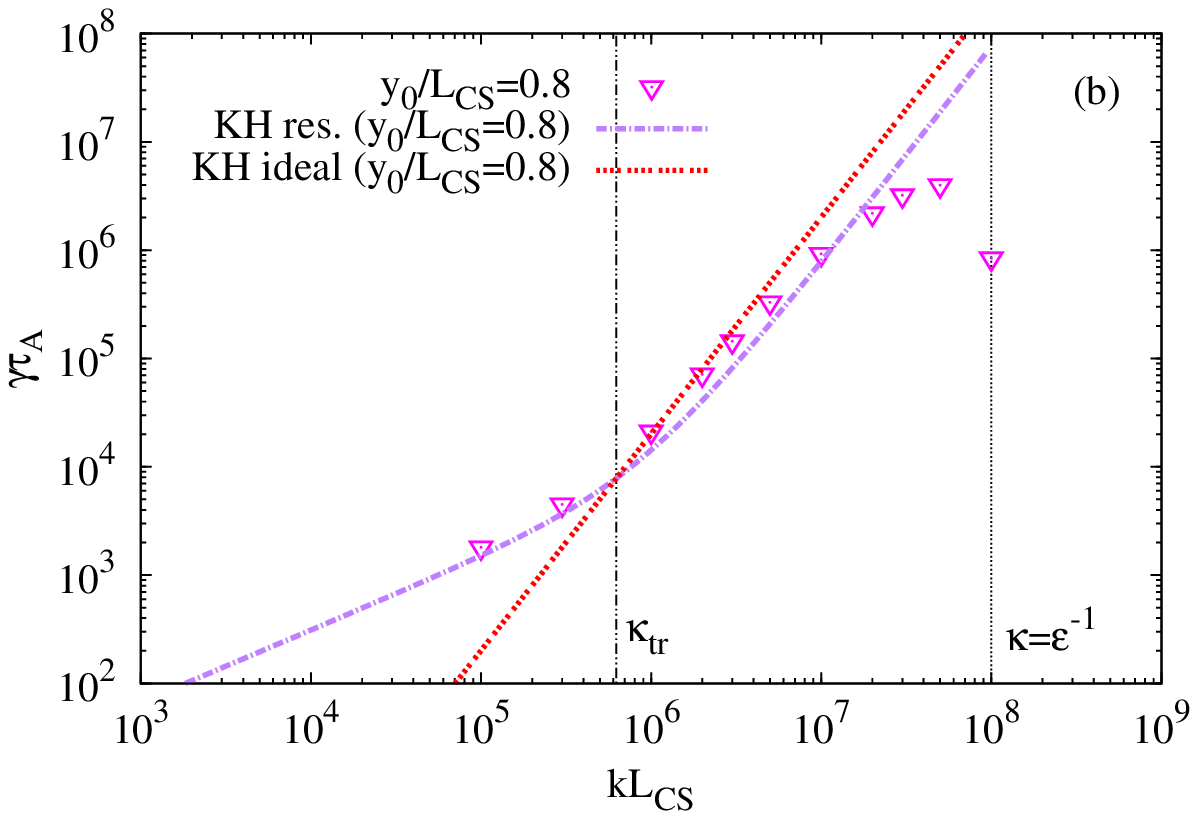}
\caption{(a) Growth rate (normalized to the global Alfv\'en 
time $\tau_A=\Lsheet/V_A$) as a 
function of the wave number $\kappa = k \Lsheet$ for
$\epsilon=10^{-6}$ (i.e., $S=10^{12}$) and at positions along the sheet, $\normyzero$.
Thin colored lines show the solution of the analytical dispersion relation for the plasmoid 
instability,~\eq{disprel}. The orange long-dash--dot line is the solution of the 
analytical resisitive KH 
dispersion relation, \eq{Lambda_1}, and the red dotted line is 
the solution of the analytical ideal KH dispersion relation, 
\eq{KH_DR_simple}, both evaluated for $\normyzero=0.8$.
Symbols are the results of direct numerical integration of the linearized equations.
The dotted black lines identify the analytically predicted slopes for the plasmoid 
instability ($\kappa^{2/3}$ for  $\kappa\lesssim\epsilon^{-3/4}$  and 
$\kappa^{-2/5}$ for $\kappa\gg\epsilon^{-3/4}$). 
The vertical dotted line is at $\kappa\epsilon=1$; the vertical dash-dot-dot line 
identifies $\kappa_{\rm tr}$ (\eq{KH_kappa_tr}).
The analytical dispersion relations are numerically solved assuming that the global equilibrium can 
be approximately described by the analytical solution in the vicinity of the origin 
derived in Ref.~[\onlinecite{uzdensky_viscous_1998}]. For this equilibrium, $\normyzerocrit=0.61$.
The numerical results use the analytical SP equilibrium calculated in 
Appendix~\ref{equilibrium} (for which $\normyzerocrit$ is the same).
(b) Same as (a) for $\normyzero=0.8$ and 
$\epsilon=10^{-8}$ (i.e., $S=10^{16}$). 
}
\label{fig:disprel}
\end{figure*}

We begin by focusing on the inviscid limit, $Pm=0$.
Plotted in~\Fig{fig:disprel}(a) are solutions of: (i) the full 
analytical dispersion relation for the plasmoid 
instability,~\eq{disprel}, for $\epsilon=10^{-6}$ (i.e., $S=10^{12}$) and 
several different values of $\normyzero$ (thin colored lines); 
(ii) the analytical KH dispersion relation in the resistive limit, \eq{Lambda_1} 
(orange, long-dash--dot line), and in the ideal limit, \eq{KH_DR_simple} (red dotted line), 
for $\normyzero=0.8$ (i.e., beyond the Alfv\'en Mach point $\normyzerocrit=0.61$). 
Overplotted (symbols) are the numerical results.
The theoretically predicted slopes for the plasmoid instability in the 
limits $\Lambda\ll 1$ and $\Lambda\rightarrow 1^-$ 
[$\kappa^{2/3}$ and 
$\kappa^{-2/5}$, respectively; see \eqs{Lambda_small}{Lambda_1}] are shown by the dotted 
black lines. The vertical dotted line is at $\kappa\epsilon=1$; 
all the analytical dispersion relations plotted are only valid for values of $\kappa$ 
significantly to the left of this line.
The vertical dash-dot-dot line shows the position of $\kappa_{\rm tr}$, the value of $\kappa$ 
at which a transition from the 
resistive to the ideal KH mode occurs, as given by \eq{KH_kappa_tr}.

\begin{figure*}[t!]
\includegraphics[width=0.48\textwidth]{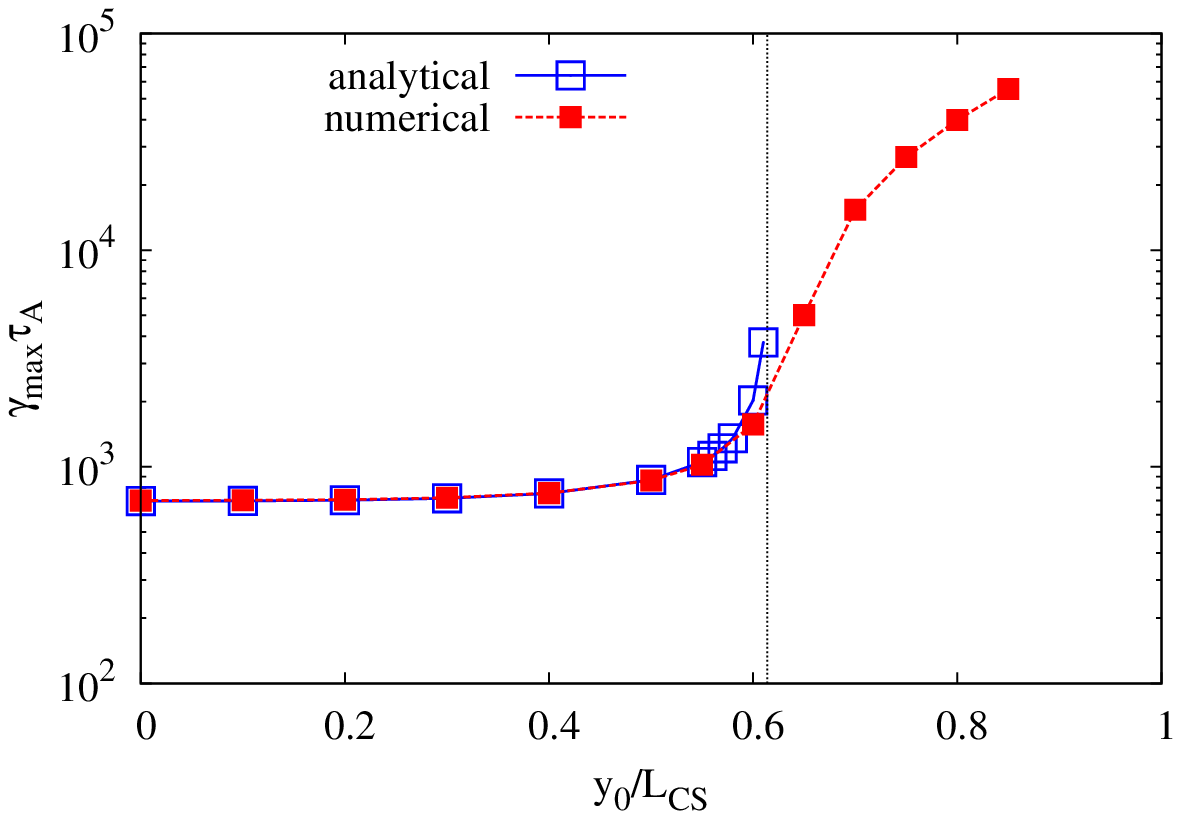}
\includegraphics[width=0.48\textwidth]{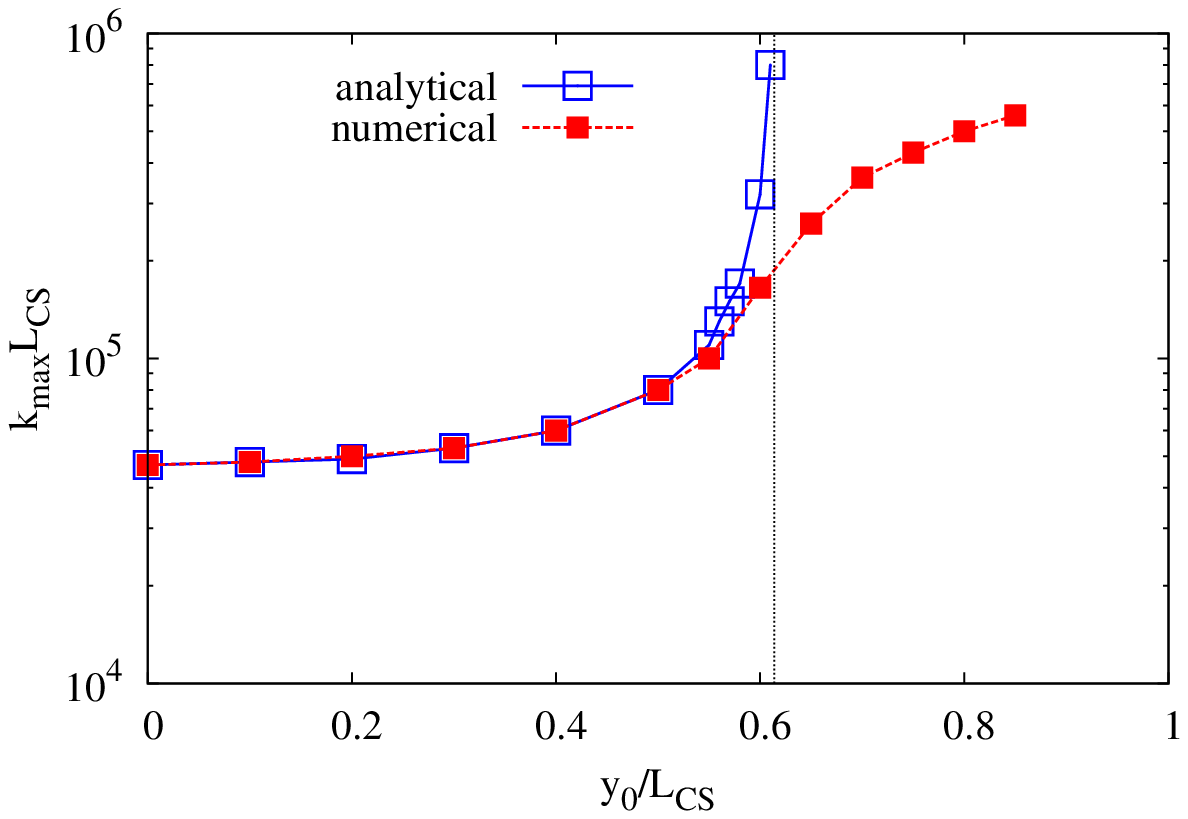}
\caption{Maximum growth rate (left) and the corresponding wave-number (right)
as functions of the position $\normyzero$ along the sheet, for
$\epsilon=10^{-6}$ (i.e., $S=10^{12}$). The analytic solution 
is given by \eqs{k_max}{gamma_max}.
The vertical dotted line identifies the 
position of $\normyzerocrit$ for the equilibrium parameters specified 
in~\eq{uz_coeffs2}.}
\label{fig:gmax}
\end{figure*}

For $\normyzero\le 0.4$, the agreement between the numerical solution 
and the analytical plasmoid dispersion relation is very good up to wave numbers 
approaching $\kappa \epsilon\sim 1$; this is not surprising since 
we used the ordering $\kappa\epsilon\ll1$ in our calculation. 
Importantly, this ordering is indeed respected by $\kappa=\kmax$, ~\eq{k_max}, 
which is accurately described 
by the analytical solution at these values of $\normyzero$.
Note that the dependence of $\gamma$ on $\normyzero$ is very weak, 
both as predicted by theory [\eq{gamma_max} with $f'_0,~v_0$ and $\finf$ given by 
Eqs.~(\ref{eq:finf}), (\ref{equilib_fprime0}) and (\ref{uz_coeffs2}), 
respectively]
and as determined by the numerical solution (see \fig{fig:gmax}).

The behavior of the growth rate is distinctly different for $\normyzero=0.8$.
This value of $\normyzero$ is above the Alfv\'en Mach point, 
$\normyzerocrit=0.61$, and it is, therefore, in the KH-unstable part of the current sheet.
At low values of $\kappa$, the analytical dispersion relation, \eq{Lambda_1} 
(labelled ``KH res.'' in the figure)
correctly captures the numerical solution. The transition from the resistive KH to 
the ideal KH mode occurs at $\kappa\approx\kappa_{\rm tr}$ given by \Eq{KH_kappa_tr}.
The ideal KH dispersion relation derived in \secref{KH}, \eq{KH_DR_simple} (labelled ``KH ideal'' 
in the figure), applies for $\kappa>\kappa_{\rm tr}$, but fails to capture the fastest growing mode,
which now occurs when $\kappa\epsilon\sim 1$, as anticipated. 

The transition between the resistive and ideal modes is not completely clear 
in \fig{fig:disprel}(a); for $\kappa>\kappa_{\rm tr}$ 
the numerical data points lie between the ideal and resistive KH lines 
(red, dotted, and orange, long-dash--dot, respectively) before rolling over and 
reaching $\gmax$ at $\kappa=\kmax$. This is because, even at such a 
large value of $S$, there is not enough scale separation between 
$\kappa_{\rm tr}\sim \epsilon^{-3/4}$, and $k_{\rm max}\sim \epsilon^{-1}$ for 
the ideal solution to match its asymptotic form derived 
in \secref{KH}.
In order to illustrate clearly this transition, we plot in \fig{fig:disprel}(b) the 
dispersion relation obtained at $\normyzero=0.8$ for an even smaller 
(perhaps unrealistically so)
$\epsilon=10^{-8}$, i.e., $S=10^{16}$.
In this figure, it is clear that the mode becomes ideal for $\kappa>\kappa_{\rm tr}$, and is 
correctly described there by the ideal KH dispersion relation, \eq{KH_DR_simple}.

\begin{figure*}
\includegraphics[width=0.48\textwidth]{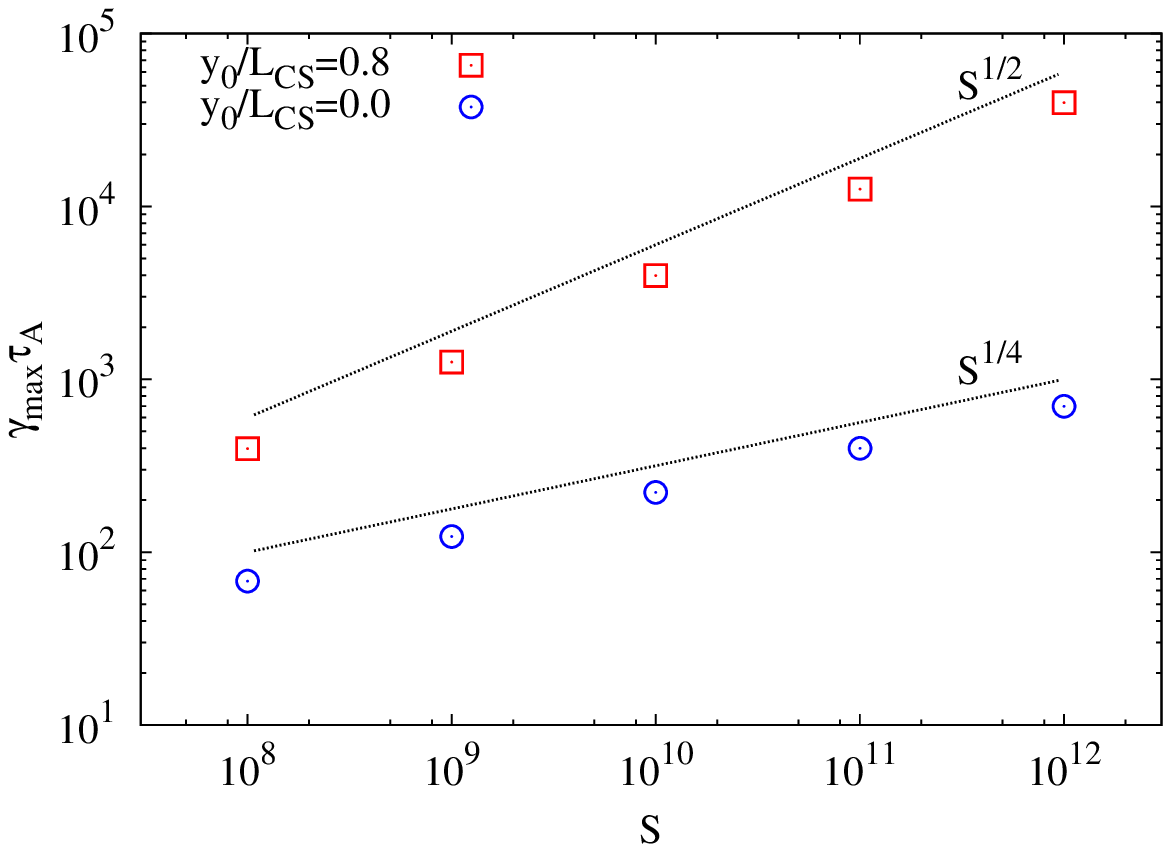}
\includegraphics[width=0.48\textwidth]{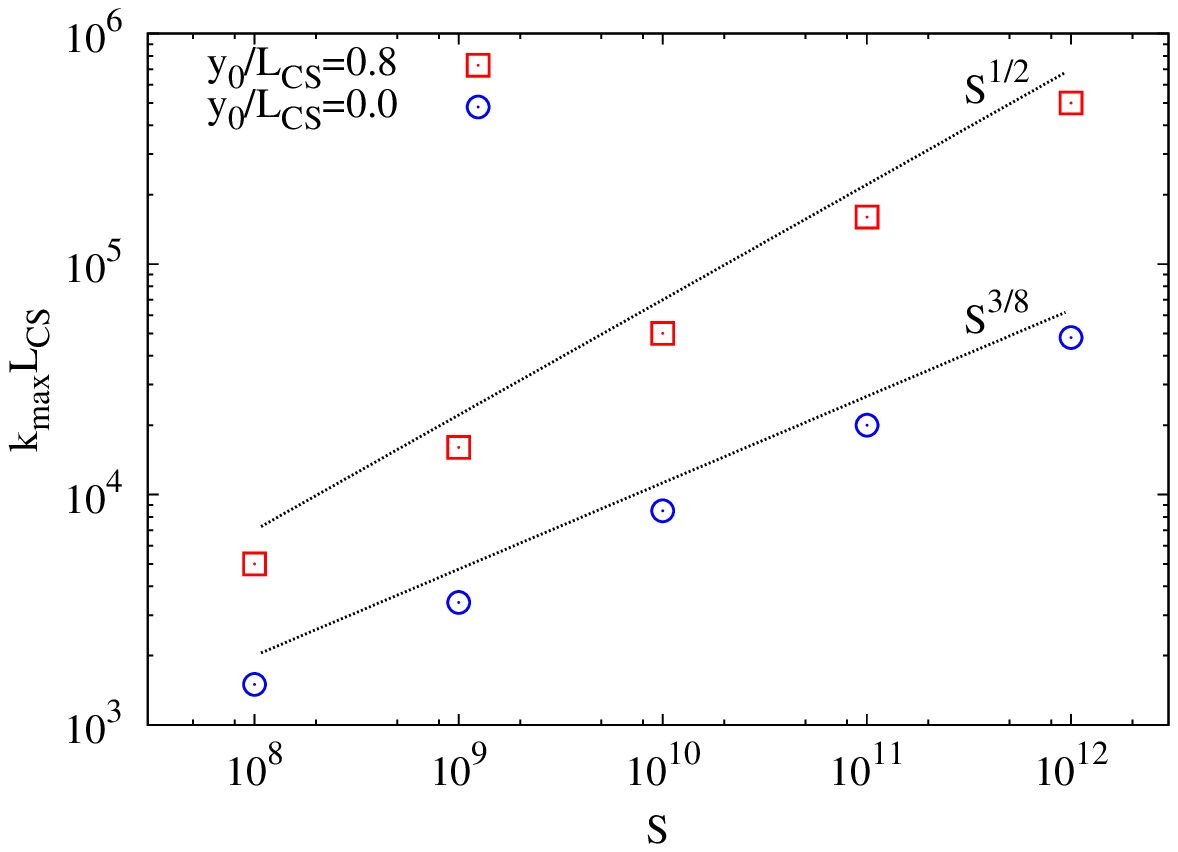}
\caption{Maximum growth rate (left) and the corresponding wave-number (right)
as functions of the Lundquist number $S$, for $\normyzero=0.0$ and $\normyzero=0.8$. 
The plasmoid instability [$\gmax\sim S^{1/4},~\kmax\sim S^{3/8}$; see 
Eqs.~(\ref{gmax_plasmoid},~\ref{gamma_max})]
observed at $\normyzero=0.0$ is superseded by the KH 
instability  [$\gmax\sim S^{1/2},~\kmax\sim S^{1/2}$; see~\eq{heur_KH_scalings}] at $\normyzero=0.8$.
}
\label{fig:gmax_vs_S}
\end{figure*}

In \fig{fig:gmax_vs_S}, we plot the 
fastest growth rate and the corresponding wave number as functions of the 
Lundquist number. Whereas the plasmoid instability scalings are obtained at $\normyzero=0.0$
(i.e., $\gmax\sim S^{1/4}$, see Eqs.~(\ref{gmax_plasmoid},~\ref{gamma_max})),
we see that it is the KH scaling that is manifest at $\normyzero=0.8$ 
(i.e., $\gmax\sim S^{1/2}$, as derived in~\eq{heur_KH_scalings}).

Plots of $\gmax$ and $\kmax$ as functions of $\normyzero$ at $S=10^{12}$ are shown in 
\fig{fig:gmax}. In this figure, the dashed vertical line identifies the position 
of the Alfv\'en Mach point, $\normyzerocrit\approx0.61$. 
As expected based on the previous discussion,
the agreement between the analytical plasmoid dispersion relation, \eq{disprel}, and 
the numerical solution is excellent at values of $\normyzero<\normyzerocrit$.
As $\normyzero\rightarrow \normyzerocrit$, the difference between the analytical and numerical 
solutions increases
and explodes  at $\normyzero=\normyzerocrit$.
The transition to the KH mode happens then; for $\normyzero\ge \normyzerocrit$, our simplified 
analytical theory (\secref{KH}) fails to produce a maximum of the growth rate and 
so cannot be compared to the numerical solution, which confirms 
that the reason for the failure of the asymptotic theory is that for $\gamma=\gmax$,
$\kmax\epsilon$ is not small, but approaches values of order unity.

\begin{figure*}
\includegraphics[width=0.48\textwidth]{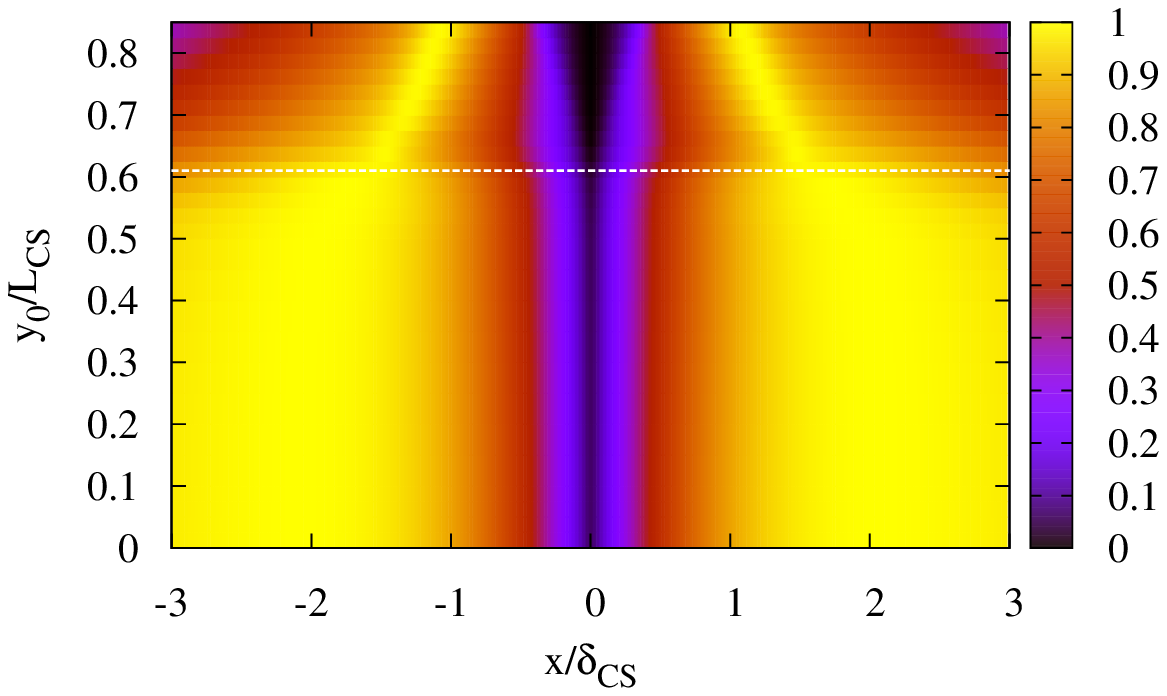}
\includegraphics[width=0.48\textwidth]{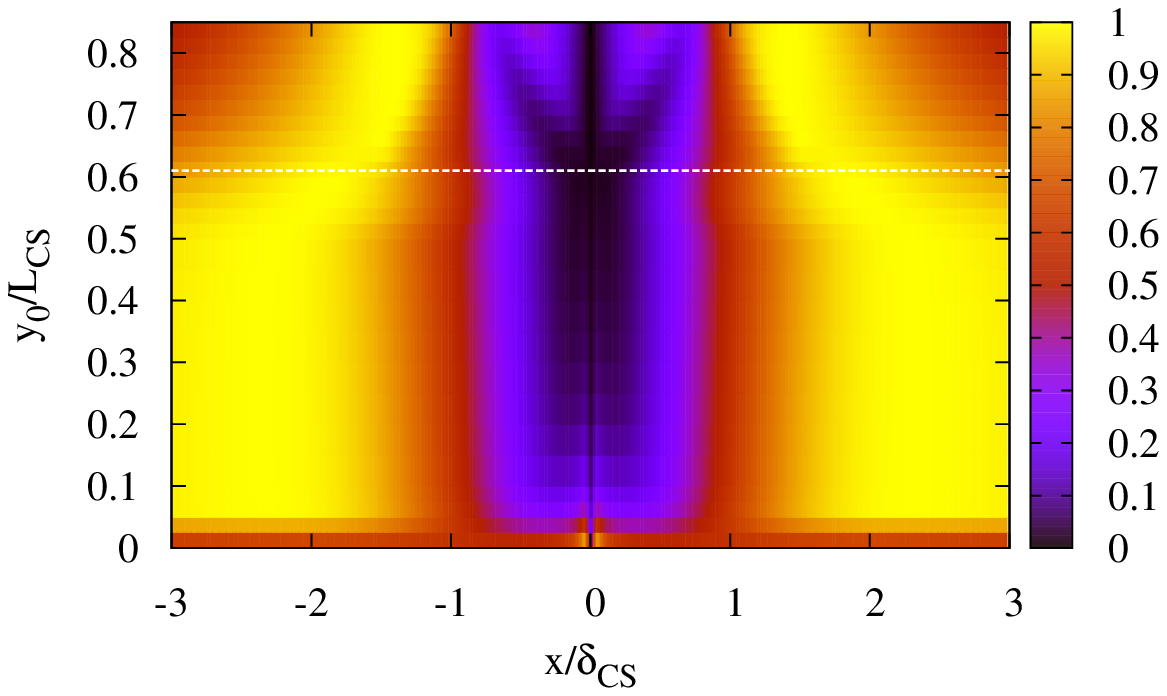}
\caption{Eigenfunctions ($|\Psi|$, left, and $|\Phi|$, right, normalized to their 
respective maxima) for $\kappa=\kmax$ at 
$\epsilon=10^{-6}$ (i.e., $S=10^{12}$). These plots are constructed from runs 
at the values of $\normyzero$ shown in \fig{fig:gmax}. In all runs, the size of the 
simulation domain is $L_x=100$, and the resolution is $\Delta x = 0.0125$.
Only a fraction of the simulation domain is shown. The horizontal dashed white line shows 
the location of the Alfv\'en Mach point, $\normyzero=\normyzerocrit$.
}
\label{fig:eigenfuncs}
\end{figure*}

\begin{figure*}
\includegraphics[width=0.48\textwidth]{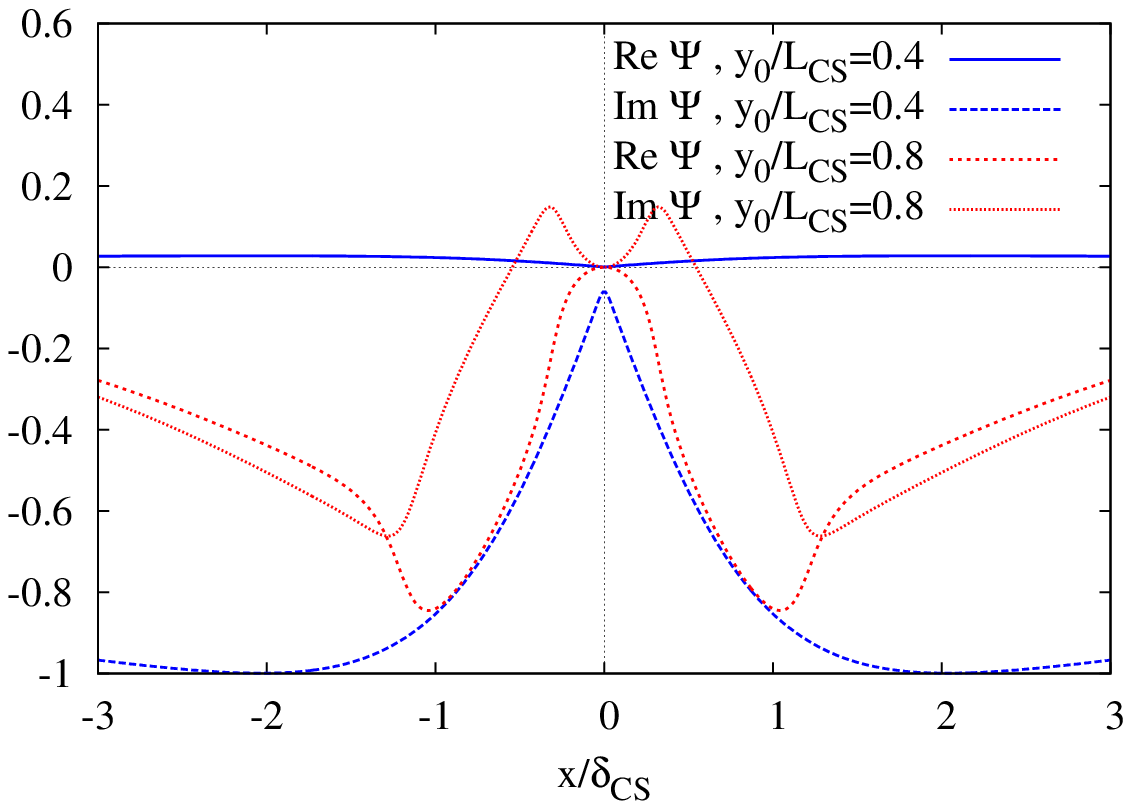}
\includegraphics[width=0.48\textwidth]{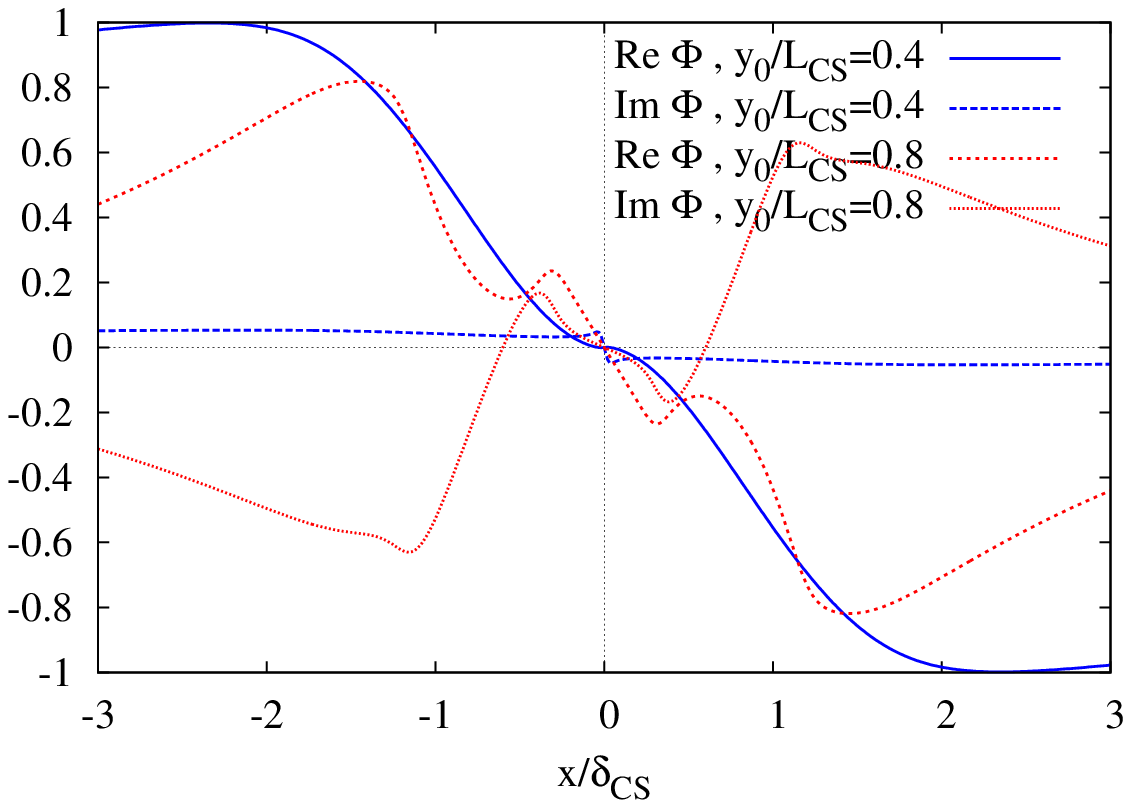}
\caption{Real and imaginary parts of the eigenfunctions 
for $\kappa=\kmax$ at 
$\epsilon=10^{-6}$ (i.e., $S=10^{12}$) and $\normyzero=0.4$ and $\normyzero=0.8$.
Only a fraction of the simulation domain is shown. 
}
\label{fig:eigenfuncs_1D}
\end{figure*}

Finally, in \fig{fig:eigenfuncs}, we show the eigenfunctions $\Psi(\xi)$ (left) 
and $\Phi(\xi)$ (right) 
corresponding 
to the fastest growing wavenumber, $\kmax$, at $S=10^{12}$. 
These plots are constructed 
from runs at the values of $\normyzero$ plotted in \fig{fig:gmax}.
We see that the eigenfunctions undergo an abrupt change at $\normyzero\sim 0.6$, 
where the Alfv\'en Mach point of the system is located (identified by the dashed white line).
Close inspection of the $\Psi$ eigenfunctions reveals that for $\normyzero>0.6$, 
$\Psi(0)\rightarrow 0$, as predicted in \secref{KH}; i.e., the most unstable mode is ideal.
Visible in the plot of the $\Phi$ eigenfunction beyond the Alfv\'en Mach point is the 
formation of structure at each of the KH-unstable shear layers, 
located at $x/\deltacs\approx \pm1$. 
For a clearer observation of both 
these properties of the eigenfunctions, we plot in \fig{fig:eigenfuncs_1D} one-dimensional 
cuts of \fig{fig:eigenfuncs}, taken at $\normyzero=0.4$ (i.e., below the Alfv\'en Mach point) 
and $\normyzero=0.8$ (i.e., above the Alfv\'en Mach point). 
As seen, for $\normyzero=0.4$, 
$\Psi$ is finite at $\xi=0$, whereas it is zero at $\normyzero=0.8$, 
in agreement with the analytical theory of  \secref{KH}  and the prediction that 
above $\normyzerocrit$ the fastest growing mode is non-reconnecting. Furthermore, 
the broadening of the $\Psi$ eigenfunction 
around $\xi=0$ at $\normyzero=0.8$ suggests the pile-up of the magnetic field that we 
discussed in \secref{KH}.


\subsection{Effect of viscosity}
\label{sec:viscosity}
In order to address the effect of viscosity on the plasmoid instability, the term 
$Pm\(\Phi''''-2\kappa^2\epsilon^2\Phi''+\kappa^4\epsilon^4\Phi\)/\kappa$, where $Pm = \nu/\eta$,
is added to the right-hand side of~\eq{IV_vort}, and two additional boundary conditions are used:
$\Phi''(-L_x,t)=\Phi''(L_x,t)=0$.
We follow the generalization of the SP 
scalings to plasmas where $Pm\gg 1$ derived in Ref.~[\onlinecite{park_reconnection_1984}]:
namely, we must scale 
the electric field at the origin as $E_0/Pm^{1/4}$,
and the width of the current layer as
$\deltacs\rightarrow\delta_{SP} Pm^{1/4}$, where $\delta_{SP}/\Lsheet=S^{-1/2}$. 
Therefore, we rescale the 
parameter $\epsilon=\deltacs/\Lsheet$ 
according to
$\epsilon\rightarrow\epsilon_{SP}Pm^{1/4}$, where $\epsilon_{SP}=\delta_{SP}/\Lsheet=S^{-1/2}$.

\begin{figure*}
\includegraphics[width=0.48\textwidth]{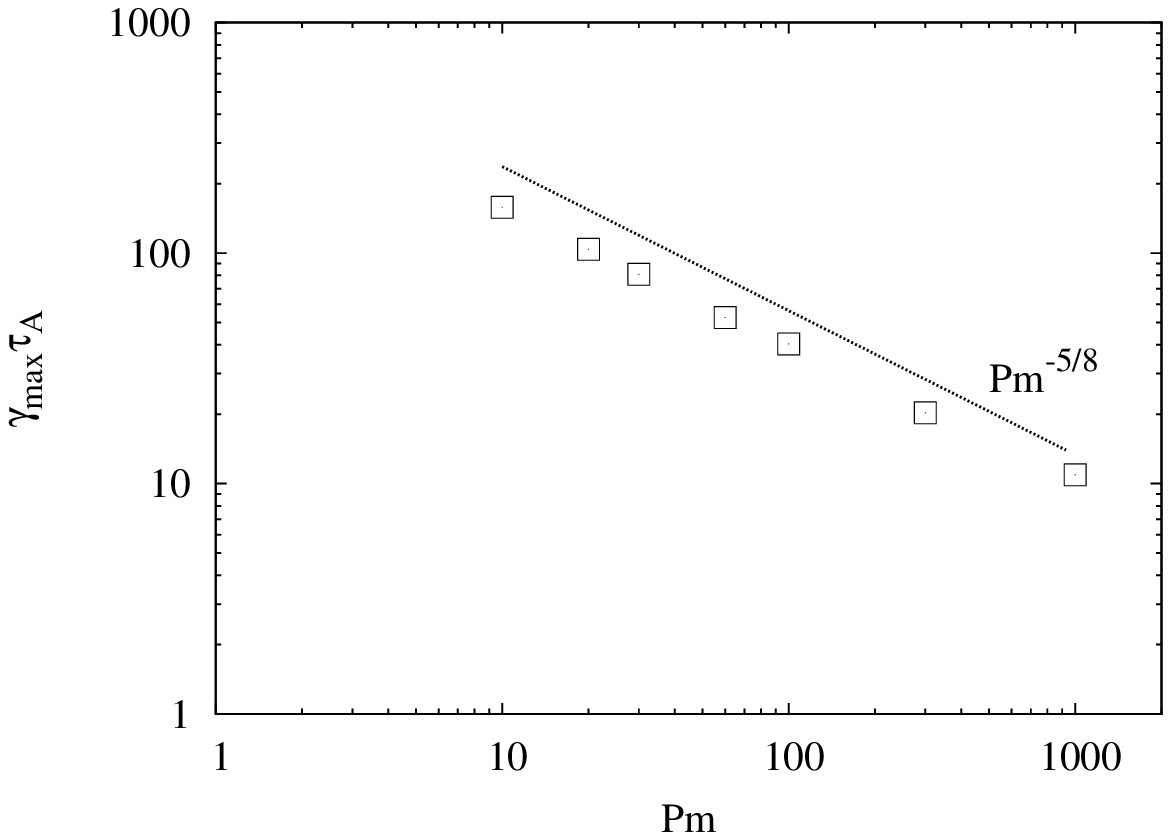}
\includegraphics[width=0.48\textwidth]{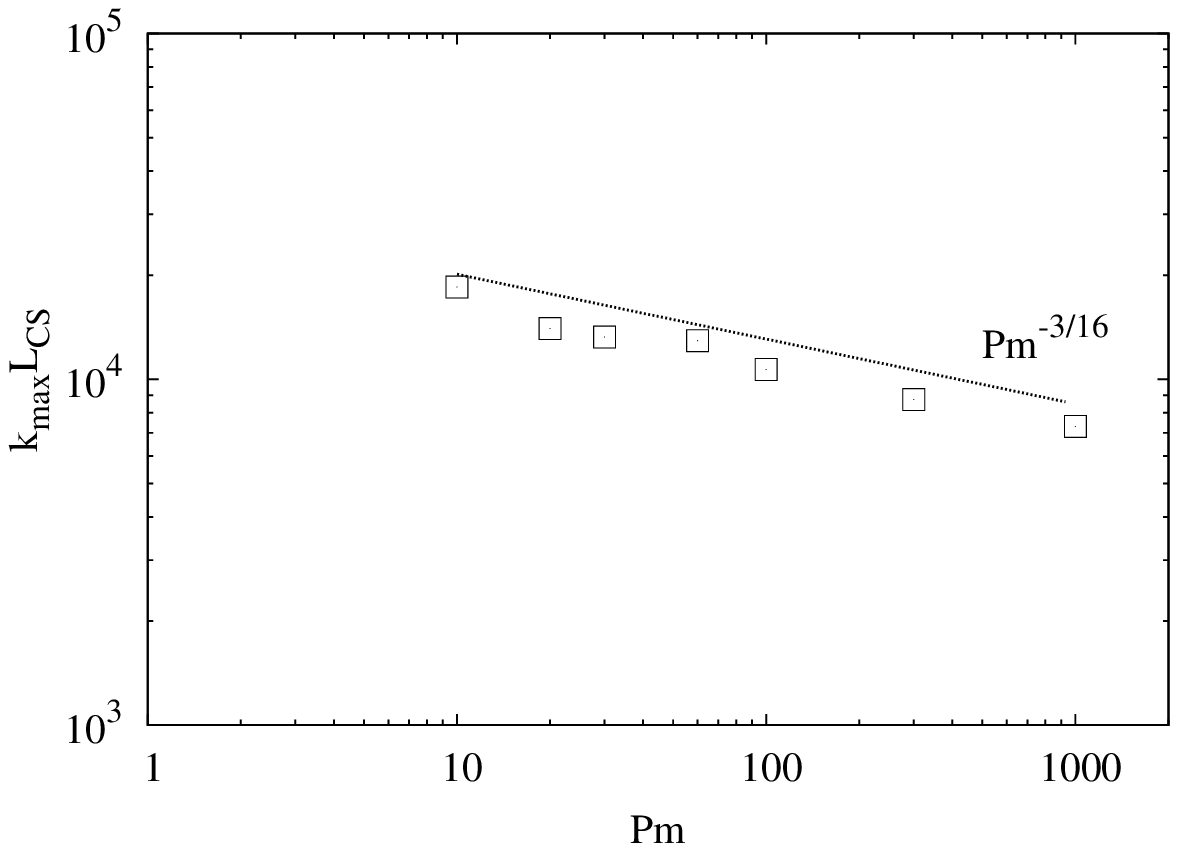}
\caption{Maximum growth rate (left) and the corresponding wave-number (right) 
as a function of the Prandtl number $Pm$ for 
$\epsilon=10^{-6}$ (i.e., $S=10^{12}$) and $\normyzero=0$.
}
\label{fig:gmax_Pm}
\end{figure*}

\begin{figure*}
\includegraphics[width=0.48\textwidth]{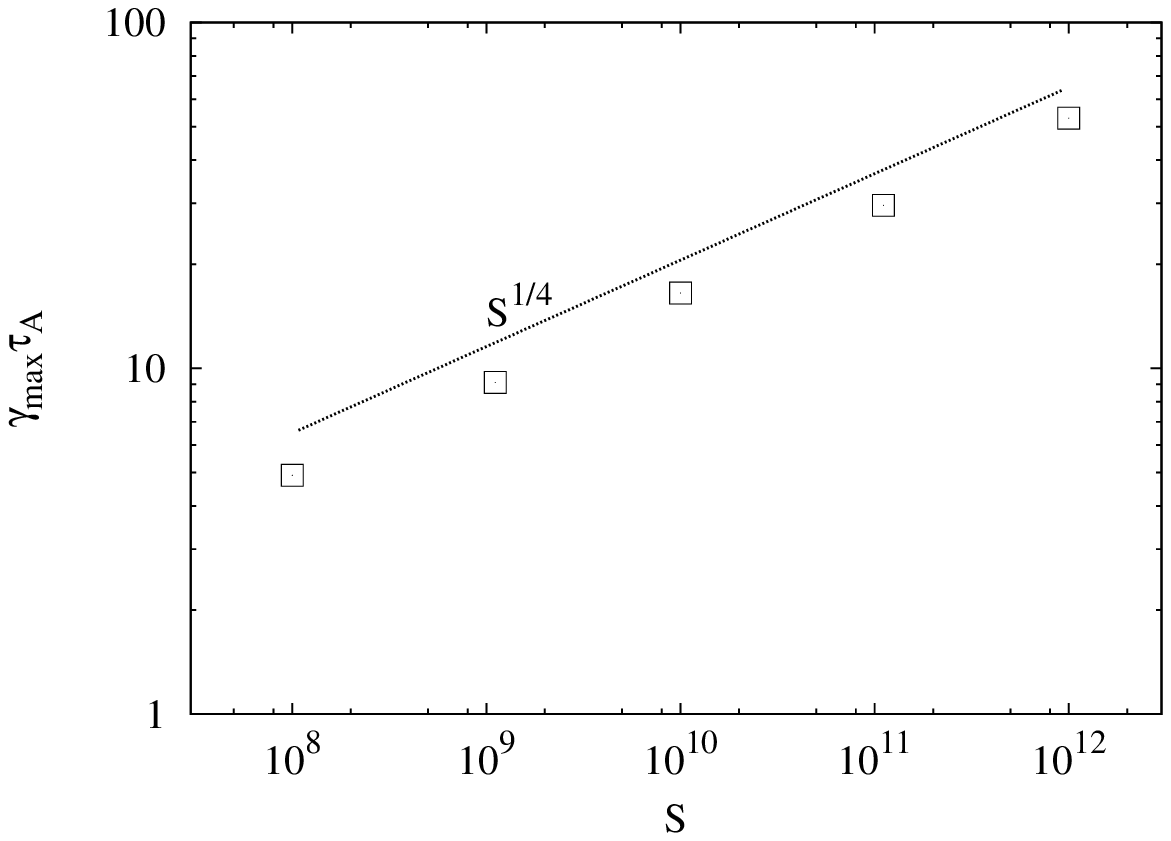}
\includegraphics[width=0.48\textwidth]{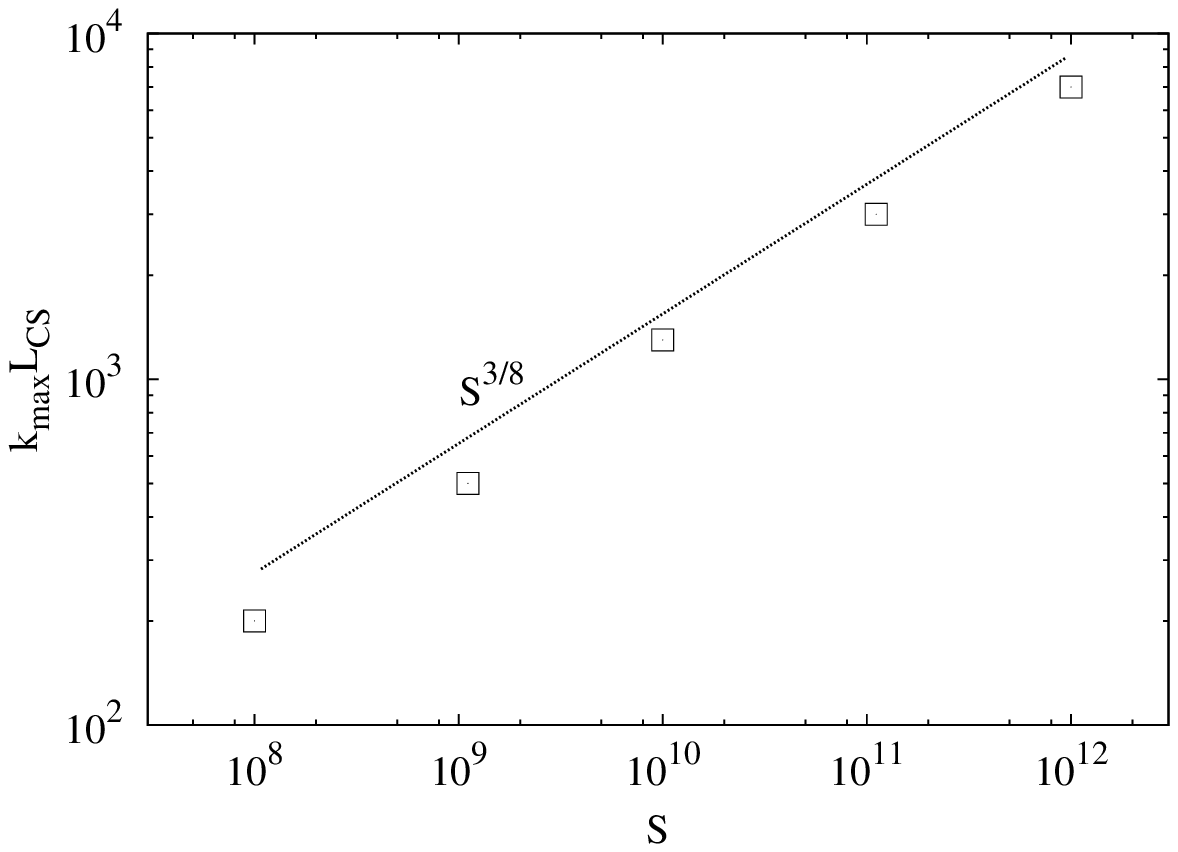}
\caption{Maximum growth rate (left) and the corresponding wave-number (right)
as a function of the Lundquist number $S$ at $Pm=30$ and $\normyzero=0$.
}
\label{fig:gmax_S_Pm30}
\end{figure*}

Plotted in~\fig{fig:gmax_Pm} are the 
maximum growth rate and the corresponding wave-number as a function of $Pm$, 
for $S=10^{12}$ and $\normyzero=0$. 
Both $\kmax$ and $\gmax$ are seen to decrease with increasing Prandtl number;
a good fit to the data is given by 
$\gmax\propto Pm^{-5/8}$ and $\kmax\propto Pm^{-3/16}$.
The scaling of $\gmax$ and $\kmax$ with the Lundquist number at $Pm=30$ is shown 
in \fig{fig:gmax_S_Pm30}. We see that the $S$ dependence of the maximum 
growth rate and of the corresponding wavenumber remains unchanged at large $Pm$, i.e., 
$\gmax\propto S^{1/4}$ and $\kmax\propto S^{3/8}$.
These results agree exactly with the power laws derived in \secref{sec:heuristic},
\eqs{kmax_largePm}{gmax_largePm}.

\section{Summary}
\label{sec:conclusions}
In this paper, a two-dimensional linear theory of the instability of 
large-aspect-ratio, Sweet-Parker-like current sheets is presented.
This work is a direct generalization of our previous 
results~\cite{loureiro_instability_2007} (Paper I), where the
simple equilibrium used was only a good model 
of a current sheet in the immediate vicinity of $y=0$ ($y$ is the outflow direction).

In the work presented here a general 2D SP-like current sheet equilibrium is considered. 
As in Paper I, we conclude that large-aspect-ratio 
Sweet--Parker current sheets are violently unstable to high-wave-number tearing-like 
perturbations,
and the same scalings of the growth rate with 
the Lundquist number $S=L V_A/\eta$ are obtained here: $\gmax \tau_A\sim S^{1/4}$ and 
$k_{\rm max} \Lsheet \sim S^{3/8}$ --- see \eqs{k_max}{gamma_max}. 
The plasmoid chain is formed inside a boundary layer whose width scales 
as $\deltain/\deltacs\sim S^{-1/8}$. These scalings have been confirmed 
via direct numerical simulation~\cite{samtaney_formation_2009,ni_linear_2010}.

The more general approach employed in this paper has allowed us to 
calculate the growth rate of the plasmoid
instability as a function of the position along the current sheet, $y_0$. 
The dependence of $\gmax$ and $k_{\rm max}$ on $y_0$ is a nontrivial function of 
the particular equilibrium 
considered and, in the absence of a known analytical solution to the 
SP problem, cannot be evaluated explicitly.
However, for $y_0/\Lsheet\ll1$ 
we make use of the semi-analytical results of Uzdensky and 
Kulsrud~\cite{uzdensky_viscous_1998} and present an exact solution --- 
Eqs.~(\ref{k_max_uz},~\ref{gamma_max_uz}). 
The most unstable
wave-number and corresponding growth rate are then found to \textit{increase} with 
distance from the center. 
Under general conditions (Syrovatskii-like upstream magnetic-field profile
and outflow profile increasing monotonically along the layer), we
show that the 
same result holds true at arbitrary $y_0/\Lsheet\sim1$.
This finding is somewhat counterintuitive:
\textit{a priori}, one could expect that the 
increasing strength of the reconnected field along the sheet, as well as the 
shear in the ouflow (in the $y$ direction), would provide a stabilising effect. 
Our calculation shows, 
however, that both are irrelevant to the instability. 
An intuitive 
understanding of why that should be so can be gained by comparing 
the strength of the upstream and the downstream magnetic fields 
at the boundary of the inner (plasmoid) layer, $\xi=\deltain$:
\be
\left. \frac{B_y}{B_x}\right|_{x\sim\deltain}\sim  \frac{\deltain}{\deltacs} S^{1/2} 
\sim S^{3/8}\gg 1,
\ee
i.e., even at the scale of the inner layer the reconnected field $B_x$ is 
completely overwhelmed by the reconnecting field $B_y$.
The gradient of the background outflow in the $y$ direction,
whose length scale is $\sim\Lsheet$, is also unimportant 
because $k_{\rm max}\Lsheet\gg1$ everywhere in the sheet.
At the periphery of the sheet,
for $y_0 > y_{0,\rm crit}$, where $y_{0,\rm crit}/\Lsheet$ is equilibrium-dependent 
but otherwise $\mathcal O(1)$, 
the current sheet becomes unstable to the Kelvin-Helmholtz (KH) instability 
driven by the velocity shear between the Alfv\'enic reconnection outflow 
and the stationary upstream plasma.
This occurs because the magnitude of the upstream magnetic field is a decreasing function 
of the outflow coordinate $y$ and eventually becomes smaller than the outflow speed (which 
is an increasing function of $y/\Lsheet$).
At, and beyond, the Alfv\'en Mach point, where this happens, 
the magnetic field can no longer stabilize the 
current sheet against the KH instability. 

We find that the KH instability of the sheet can either be resistive 
(i.e., induce reconnection at $x=0$), 
or ideal (no reconnection), with lower values of $k\Lsheet$ corresponding to the former, 
and larger values to the latter.
The fastest growing KH mode, $k_{\rm max}\Lsheet\sim S^{1/2}$ (i.e., $k_{\rm max}\deltacs\sim 1$), 
is an ideal, non-reconnecting mode.
This is because reconnection cannot occur at the fast rates 
required by the fastest growing KH mode. 
A useful analogy can be made with the Taylor (forced reconnection) problem~\cite{hahm_forced_1985}: 
since there are two shear layers, one on each side of 
the current sheet, 
the KH instability of the sheet is conceptually similar to a situation where perturbations at distant walls
attempt to drive  reconnection at a rational surface.  
In the Taylor problem, it is also found that 
the perturbations at the walls do not drive reconnection in the initial stage.
However, the same analogy suggests that as the ideal KH mode evolves 
into the nonlinear regime, it will cause the upstream magnetic field 
to pile-up in the current layer, eventually leading to its reconnection.
This KH-driven reconnection that occurs at  $y_0 > y_{0,\rm crit}$ will give rise 
to a plasmoid chain, just as the ``pure'' plasmoid instability that is found at 
 $y_0 < y_{0,\rm crit}$.
Therefore, in practice, it may be difficult to distinguish between the two situations. 

It is also worth noting that the basic KH instability mechanism that we have described here is
completely general, i.e., it should apply 
to {\it any} reconnecting current sheet, not just to those that can be described by the reduced MHD 
framework that we have adopted here: the only ingredient it requires is the existence of an 
Alfv\'enic Mach point somewhere along the layer. 
This should be a generic feature of most reconnecting current sheets. 
Whether the layer is collisional or collisionless 
may affect the dynamics of the KH instability, but its existence is not dependent on the 
plasma collisionality. In this respect, our findings may be related to 
recent observations of the KH instability in collisionless simulations 
of guide-field reconnection~\cite{fermo_2012}.

Finally, the effect of viscosity on the plasmoid instability has been addressed 
via numerical integration of the linearised set of equations.
Our results are that in the limit $Pm = \nu/\eta \gg 1$, the fastest growth rate and wave-number
of the plasmoid instability scale as:
\be
\label{gmax_largePm_2}
\gmax\sim S^{1/4}Pm^{-5/8}\sim \Lsheet^{1/4}~V_A^{1/4} \eta^{3/8} \nu^{-5/8},
\ee
\be
\label{kmax_largePm_2}
\kmax\sim S^{3/8}Pm^{-3/16}\sim \Lsheet^{3/8}~V_A^{3/8} \eta^{-3/16} \nu^{-3/16}.
\ee
We have not performed a rigorous analytical calculation of the plasmoid 
instability in this limit, but we have been able to recover these scalings  
in a non-rigorous way from known results on the visco-tearing and 
resistive-kink modes~\cite{porcelli_viscous_1987}, via the rescaling of the background 
magnetic shear length $a\rightarrow \deltacs\sim \Lsheet S^{-1/2}Pm^{1/4}$~\cite{park_reconnection_1984}.
Although these scalings are only expected to apply for $S\gg \Scrit,~Pm\gg1$,
where $\Scrit$ is the critical value of the Lundquist number for the current sheet to be 
plasmoid-unstable, they lead to the prediction that
\be
\label{Scrit_largePm}
\Scrit\sim 10^4 Pm^{1/2}, \qquad Pm\gg1.
\ee
This result, as well as those of \eqs{gmax_largePm_2}{kmax_largePm_2} are concrete 
predictions that can be tested in direct numerical simulations of MHD reconnection 
in the large magnetic Prandtl number regime.

\section*{Acknowledgments}
The authors would like to thank S.~C. Cowley, R. Samtaney and J.~B.~Taylor 
for important discussions.
The work of N.F.L.\ was supported by Funda\c{c}\~ao para a Ci\^encia e a 
Tecnologia (Ci\^encia 2008 and Grant no. PTDC/FIS/118187/2010) 
and by the European Communities under the contract of Association 
between EURATOM and IST.
The views and opinions expressed herein do not necessarily 
reflect those of the European Comission. 
A.A.S.\ was supported in part by an STFC Advanced Fellowship and 
the STFC Grant ST/F002505/2. 
D.A.U.\ was supported by NSF Grant PHY-0903851.
N.F.L. and D.A.U. thank the Leverhulme 
Trust International Network for Magnetized Plasma Turbulence for travel 
support.
Numerical work was carried out at the IST cluster (Instituto Superior T\'ecnico), 
Newhydra (University of Oxford) and 
Verus (University of Colorado).
\appendix
\section{Equilibrium Considerations}
\label{equilibrium}
Exact two-dimensional solutions of 
\eqs{RMHD_vort}{RMHD_psi} describing a 
Sweet-Parker-like reconnecting current sheet are 
not known. 
In principle, these can be obtained by 
substituting the expressions for $\psi$ and $\phi$ given by 
\eqs{psi_eq}{phi_eq} into \eqs{RMHD_vort}{RMHD_psi} and 
equating equal powers of $(y-y_0)/\Lsheet=\bar y-\normyzero$.
To lowest order in $\bar y-\normyzero$, we obtain the following equations:
\bea
\label{ohm_lowest_order}
u(\xi)f(\xi)-\normyzero^2 v(\xi)g(\xi)&=&f'(\xi)-\bar E_0,\\
\label{vort_lowest_order}
u(\xi)v''(\xi)-v(\xi)u''(\xi)&=&g(\xi)f''(\xi)-f(\xi)g''(\xi) +\nonumber\\
&&{Pm}~v'''(\xi),
\eea 
where we have used the normalizations of \eq{normalizations}, neglected terms of 
order $\epsilon^2$, and defined the normalized electric field 
$\bar E_0=\Lsheet E_0/(B_0^2\deltacs)$.

Evaluated at $\xi=0$, \eq{ohm_lowest_order} yields:
\be
\label{equilib_fprime0}
f'_0=\bar E_0-\normyzero^2v_0g_0,
\ee
whereas for $\xi\gg1$ we obtain from the same equation
\be
\uinf=\frac{\bar E_0}{\finf}
\ee
These expressions are exact; however, we see that \eqs{ohm_lowest_order}{vort_lowest_order}
are not a closed set, since there are only two equations and four unknowns: $f(\xi),~g(\xi)$ 
(the normalized reconnecting and reconnected magnetic field profiles, respectively), 
and $u(\xi),~v(\xi)$ (the normalized inflow and outflow velocity profiles, respectively).
This closure problem is introduced by the expansion in $(\bar y-\normyzero)$ (recall the 
discussion of \secref{sec:prob_setup}).
In order to obtain a SP-like equilibrium, which we require for our numerical solution, 
one has to close \eqs{ohm_lowest_order}{vort_lowest_order}, e.g., by guessing two of 
the four unknown
functions, and solving those equations for the other two.
Any model of the equilibrium that can be found in this way 
is necessarily non-unique (i.e., dependent on the guesses required for the closure); 
however, we will see in what follows that 
a qualitatively satisfactory model of a SP current sheet can be obtained by this procedure.

Let us introduce an auxiliary function, $s(\xi)$, 
defined by the following equation:
\be
\label{eq_g_s}
g(\xi)= \frac{\uinf}{\finf}v(\xi)-s(\xi).
\ee
Then, from~\eq{ohm_lowest_order} we obtain
\be
\label{SP_v}
\begin{split}
v(\xi)= &\frac{\finf}{\uinf}\frac{s(\xi)}{2} \\
&\pm 
\sqrt{\frac{\finf^2}{\uinf^2}\frac{s^2(\xi)}{4} + 
\frac{\finf}{\uinf}\frac{u(\xi)f(\xi)-f'(\xi)+\bar E_0}{\normyzero^2}}.
\end{split}
\ee

\Eq{vort_lowest_order} can also be easily solved  in the limit 
$Pm=0$ (viscous effects 
in the equilibrium that we are about to derive can be modelled by a rescaling of 
the current sheet thickness, the outflow speed and the reconnection electric field according to the 
SP relationships in the viscous regime derived in~\cite{park_reconnection_1984}).
Using \eq{eq_g_s}, \eq{vort_lowest_order} becomes
\be
s''(\xi)=s(\xi) \frac{f''(\xi)}{f(\xi)},
\ee
to be solved subject to the boundary conditions $s(0)=\uinf v_0/\finf-g_0$, 
where $v_0=v(0)$ and $g_0=g(0)$, 
and $s'(0)=0$ (we demand that both $v(\xi)$ and $g(\xi)$ are even functions).

The general solution to this equation is
\be
\label{eq_s}
s(\xi)= C_1 f(\xi) + C_2 f(\xi) \int^\xi \frac{d \xi'}{f^2(\xi')}.
\ee
[The lower limit of integration on the last term on the right-hand side 
of this expression need not be specified as $C_1$ 
can be redefined to absorb the difference between different lower limits; note however 
that we take the lower limit to be finite, i.e., neither $0$ nor $\infty$.]

At this stage, the equilibrium problem is solved if we provide functional forms for the 
reconnecting magnetic field, $f(\xi)$, and for the inflow velocity profile, $u(\xi)$. 
The simplest choice for $f(\xi)$ is the ``Harris sheet''~\cite{harris_1962}:
\be
\label{SP_f}
f(\xi)=\finf\tanh\(\frac{f'_0}{\finf} \xi\).
\ee
A qualitatively plausible choice for $u(\xi)$ is
\be
\label{SP_u}
u(\xi)=-\uinf\frac{f(\xi)}{\finf}.
\ee
Substituting \eq{SP_f} into \eq{eq_s} and evaluating the integral explicitly, we obtain:
\be
\label{SP_s}
s(\xi)=\(g_0-\frac{\uinf}{\finf}v_0\)
\[\frac{f'_0}{\finf}\xi\tanh\(\frac{f'_0}{\finf}\xi\)-1\],
\ee
where the constants of integration $C_1,~C_2$ have been chosen 
to satisfy the boundary conditions we specified for $s(\xi)$. 
Substituting \eqs{SP_f}{SP_s} into \eqs{eq_g_s}{SP_v} yields 
explicit expressions for the two remaining unknowns, the reconnected magnetic 
field $g(\xi)$ and 
the outflow velocity profile, $v(\xi)$.
Although it is not particularly enlightening to write down these expressions in explicit form, 
it is useful to evaluate $g(\xi)$ for $\xi\gg1$. It is
\be
\label{g_at_inf}
g(\xi)|_{\xi\gg1}\approx \pm\(\frac{\uinf}{\finf}v_0-g_0\)\frac{f'_0}{\finf}\xi
\equiv \pm g_\infty'\xi.
\ee
This expression is used in \secref{ext_reg}, to estimate the magnitude of $g(\xi)$ for $\xi\gg1$. 

The last step in obtaining an analytical SP-like equilibrium solution consists of determining  
$E_0, ~v_0, ~g_0$ and $\finf$, all of which can in principle be functions of $\normyzero$. 
A reasonable choice for $\finf$ is a Syrovatskii-like profile~\cite{syrovatskii_formation_1971}:
\be
\label{eq:finf}
\finf=\sqrt{1-\normyzero^2}.
\ee
As for $E_0, ~v_0, ~g_0$, their values at $\normyzero=0$ have been calculated semi-analytically 
in Ref.~[\onlinecite{uzdensky_viscous_1998}]~\footnote{
These are the  values of the coefficients obtained numerically
in Ref.~\cite{uzdensky_viscous_1998} for the smallest value of the magnetic Prandtl 
number employed in that 
paper, $Pm=0.005$, and are assumed by us to be the converged values 
in the asymptotic case $Pm\ll 1$.}:
\begin{align}
\label{uz_coeffs2}
\bar E_0(\normyzero=0)=1.075; \nonumber\\ 
g_0(\normyzero=0)=0.642; \nonumber\\
\quad v_0(\normyzero=0)=1.286.
\end{align}
The simplest choice is to assume that these values are constant along the sheet [note, though, 
that a linearly increasing dependence of the outflow and reconnected field profiles is 
already included in the normalizations, \eq{normalizations}].

\begin{figure*}
  \includegraphics[width=0.45\textwidth]{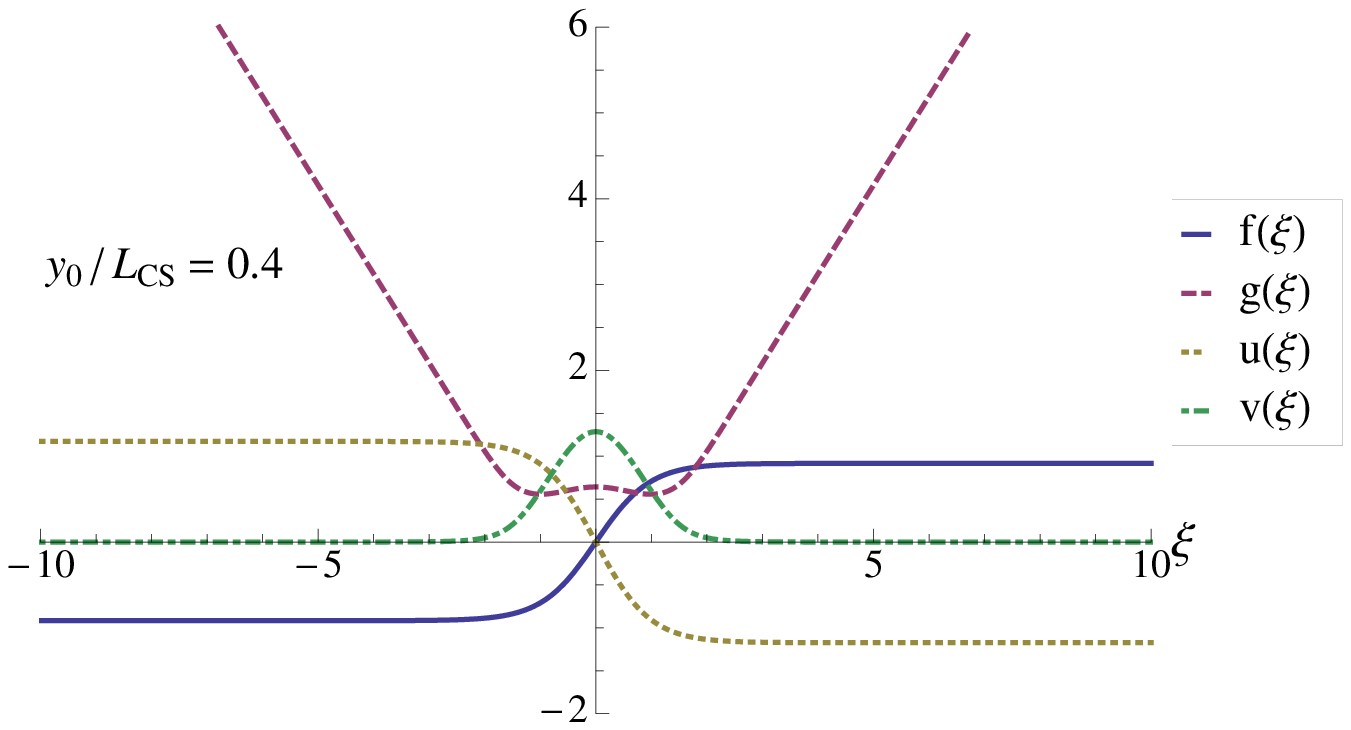}
  \includegraphics[width=0.45\textwidth]{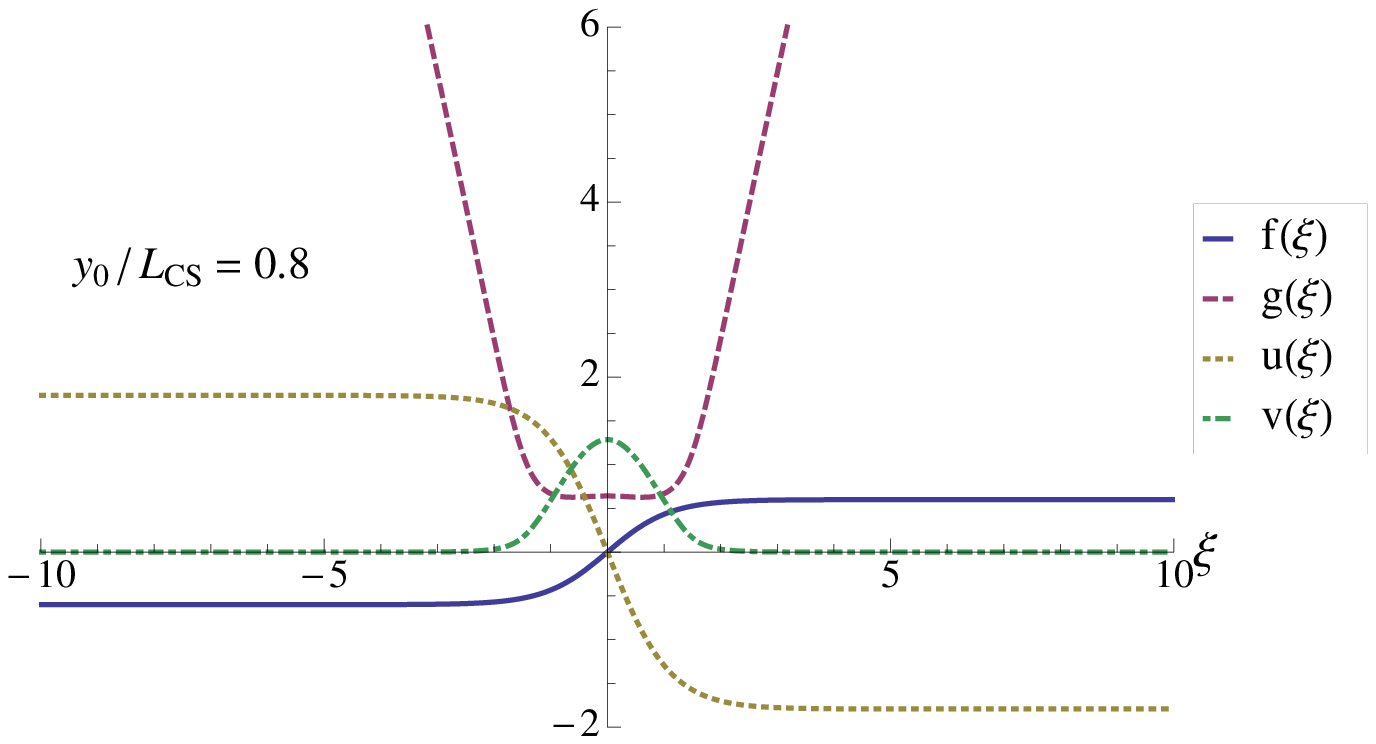}
  \caption{Analytic SP-like equilibrium profiles 
(Eqs.~(\ref{eq_g_s}),~(\ref{SP_v}),~(\ref{SP_f}),~(\ref{SP_u})) evaluated for $\normyzero=0.4$ (left), 
and $\normyzero=0.8$ (right). 
These equilibria are obtained by chosing the functional form of the upstream magnetic field, $f(\xi)$, and 
imposing that the inflow profile be such that $u(\xi)=-\uinf f(\xi)/\finf$. 
The lowest order (in $(y-y_0)/\Lsheet$) Ohm's law and momentum equation can then be solved 
for the two remaining unknowns, namely the reconnected magnetic field, $g(\xi)$, and the outflow, $v(\xi)$.}
  \label{fig:SP_equilib}
\end{figure*}

Examples of the equilibrium profiles obtained in this fashion are shown in \fig{fig:SP_equilib}, 
for $\normyzero=0.4$ (left) and $\normyzero=0.8$ (right)
(in \eq{SP_v}, the solution 
with the `$+$' sign is chosen).
We see that these profiles retain all the qualitative features expected of a true SP equilibrium.
Note that for these parameters, the 
Alfv\'en Mach point of the system occurs at $\normyzerocrit=0.61$.

The solution found here can be viewed as a generalization to the entire current sheet of 
the equilibrium derived by Biskamp~\cite{biskamp_nonlinear_1993}, which is 
only applicable for $\normyzero=0$.
As mentioned above, the equilibrium profiles obtained by this procedure, though exact, are not unique,
since they depend on the guesses for $f(\xi),~u(\xi)$; another {\it ansatz} can, in principle,
yield a different, but equally plausible, equilibrium. For the purposes of this paper, however, 
we do not believe this to be a serious constraint since we expect both 
the plasmoid and the KH instabilities to be largely independent of the fine 
details of the background profiles; this certainly seems to be true for the plasmoid instability, 
as is suggested by the agreement between the theoretical 
predictions of Paper I using a very simplified equilibrium and
subsequent numerical studies~\cite{samtaney_formation_2009, huang_scaling_2010}.
The profiles we have derived are a convenient model for solving the linear problem, as we 
do in \secref{numerics}.

%
\end{document}